%% file: acl_latex.tex
\documentclass[11pt]{article}

\usepackage[preprint]{acl}

\usepackage{times}
\usepackage{latexsym}

\usepackage[T1]{fontenc}

\usepackage[utf8]{inputenc}

\usepackage{microtype}

\usepackage{inconsolata}

\usepackage{graphicx}

\usepackage{amssymb}
\usepackage{graphicx} 
\usepackage{subcaption} 
\usepackage{algorithm}
\usepackage{algorithmicx}
\usepackage{algpseudocode}
\usepackage{booktabs}
\usepackage{colortbl}
\usepackage{caption}
\usepackage{amsmath}
\usepackage{tabularx}
\usepackage{multirow}
\usepackage{makecell}

\definecolor{mygreen}{RGB}{0,160,0}

\definecolor{myred}{RGB}{178,34,34}

\title{PaperRegister: Boosting Flexible-grained Paper Search via \\Hierarchical Register Indexing}

\author{
Zhuoqun Li${}^{1,2}$,
Xuanang Chen${}^{1}$,
Hongyu Lin${}^{1}$,
Yaojie Lu${}^{1}$, \\
\textbf{Xianpei Han${}^{1}$,}
\textbf{Shanshan Jiang${}^{3}$,}
\textbf{Bin Dong${}^{3}$,}
\textbf{Le Sun${}^{1}$} \\
${}^{1}$Chinese Information Processing Laboratory,
Institute of Software, Chinese Academy of Sciences\\
${}^{2}$University of Chinese Academy of Sciences \\
${}^{3}$Ricoh Software Research Center Beijing, Ricoh Company, Ltd. \\
{\tt \{lizhuoqun2021,chenxuanang,hongyu,luyaojie,xianpei,sunle\}@iscas.ac.cn} \\
}

\begin{document}
\maketitle

\setlength\abovedisplayskip{5pt}
\setlength\belowdisplayskip{5pt}

\begin{abstract}
As researchers delve more deeply into their work, paper search requirements may become more flexible, sometimes involving specific details such as module configuration rather than being limited to coarse-grained topics. 
However, previous paper search systems are unable to meet these flexible-grained  requirements, as previous systems mainly collect paper abstract to construct corpus index, which lacks detailed information to support retrieval by some finer-grained queries.
In this work, we propose PaperRegister, which transforms traditional abstract-based index into a hierarchical index tree, thereby supporting queries at  flexible granularity. 
Experiments on paper search tasks across a range of granularity demonstrate that PaperRegister achieves the SOTA performance, and particularly excels in the fine-grained scenarios, highlighting good potential as an effective solution for flexible-grained paper search in real-world applications.
\url{https://github.com/Li-Z-Q/PaperRegister}.
\end{abstract}

\input{sections/intro}
\input{sections/register}

\input{sections/recognizer}
\input{sections/experiment}

\input{sections/related}

\input{sections/conclusion}

\bibliography{custom}

\appendix


\input{sections/app}

\end{document}

%% file: sections/intro.tex
\section{Introduction}

Paper search is an important and almost everyday activity for researchers~\cite{kuhlthau1991inside,ellis1993comparison,hemminger2007information,case2016looking}. 
Typically, this process begins when user submits a natural-language query describing a topic, and then retrieval system matches this query against
 paper corpus and returns subset of papers with the highest relevance~\cite{wadden2020fact,cohan2020specter,ajith2024litsearch,he2025pasa}.
As researchers delve more deeply into their work, paper search requirements can become increasingly flexible. 
For example, the view of query may refer to detailed module configuration or methodological operation, rather than being limited to the level of coarse-grained topic~\cite{mysore2021csfcube,kang2024taxonomy, wang2023scientific,zhang2025scientific}. 
Therefore, a paper search system supporting queries at flexible granularity is of important value.

\input{figs/head}

Unfortunately, existing paper search systems cannot effectively handle queries across flexible granularity, since they mainly use paper abstract to construct   corpus index for retrieval~\cite{zheng2020bert,gao2023precise,mackie2023generative,lei2024corpus}. 
When the view of query involves finer‐grained information that does not appear in abstract, they fail to retrieve relevant papers. 
As shown in Figure~\ref{fig:head}, the  query \textit{``Retrieve papers about jointly training encoder and generator via minimizing negative marginal log-likelihood without doc encoder''} focuses on a detailed training operation. 
For this query, the target papers cannot be successfully retrieved by traditional abstract-based index, because paper abstract does not mention such a detailed training operation.
Therefore, building a paper search system supporting queries at flexible granularity remains a valuable and unresolved challenge.

To this end, we propose \textbf{PaperRegister}, which transforms traditional abstract-based index into a hierarchical index tree, thereby supporting queries at flexible granularity.
Specially, PaperRegister includes offline hierarchical indexing and online adaptive retrieval.
In the offline stage, to construct the hierarchical index tree for paper corpus, PaperRegister uses a hierarchical register schema, which is consisted of information nodes at various granularity, and each node path represents a kind of view. 
Based on this schema, PaperRegister first extracts fine‐grained node contents, then aggregates node contents bottom‐up, layer by layer, yielding hierarchical register for each paper. 
And then the hierarchical index tree is constructed by merging  register of all papers.
In the online stage, to adaptively retrieve via the hierarchical index tree, PaperRegister first employs a view recognizer to identify the views of query, which associates to suitable indexes from the  tree, and then matches query against these indexes to achieve the precise paper search.

One key aspect of PaperRegister is the view recognizer, which must be both low-latency and high-accuracy. 
To this end, we construct the view recognizer by training a small-scale language model via  hierarchical-reward reinforcement learning.
Firstly, to alleviate online latency, we adopt a language model with 0.6 billion parameters as base. 
Then, to equip the model with basic identifying ability, we perform supervised fine-tuning, where query is the input and golden view is the label.
Furthermore, to enhance accuracy of view recognizer, considering the hierarchical dependency of nodes in  register schema, we design a hierarchical reward as  reinforcement learning signal, which is calculated by the closeness level between  predicted view and golden view in hierarchical register schema. 
And then we employ this reward on group relative policy optimization (GRPO) ~\cite{Shao2024DeepSeekMathPT} to enhance the capability of view recognizer.

\input{figs/method}

In experiments, we compare PaperRegister with several common-used methods on paper search, across various levels of granularity. 
Results show PaperRegister achieves the SOTA performance, with the improvement being more pronounced as query granularity is finer, confirming that PaperRegister is a robust solution for the challenging flexible-grained paper search task.
The main contributions of this work can be summarized as follows:
\begin{itemize}
\setlength{\itemsep}{2pt}
\setlength{\parsep}{2pt}
\setlength{\parskip}{2pt}
\setlength{\topsep}{0pt}

\item We construct PaperRegister, which can support paper search queries at flexible granularity via the hierarchical register indexing.

\item We design hierarchical-reward reinforcement learning, which can train a powerful and low-latency view recognizer for PaperRegister.

\item We conduct extensive experiments and analysis, which show PaperRegister is an advanced system for flexible-grained paper search.
\end{itemize}

%% file: figs/head.tex
\begin{figure}[t!]
\centering
\includegraphics[width=0.99\linewidth]{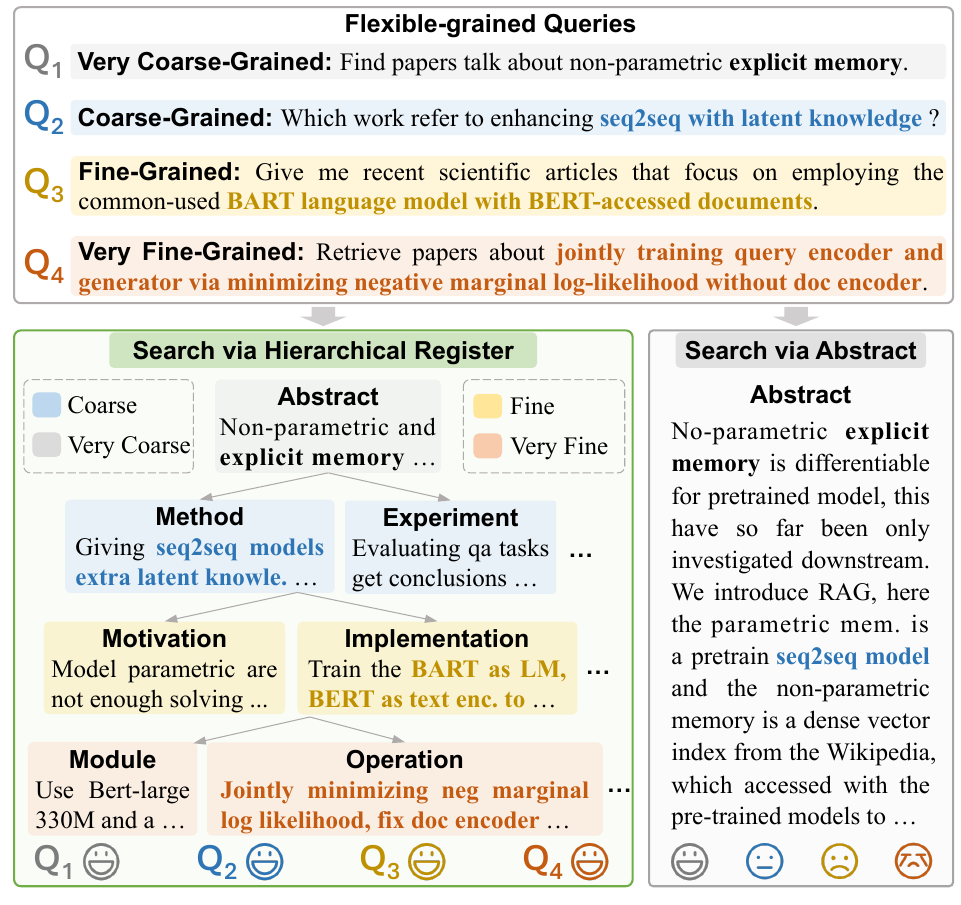} 
\caption{
PaperRegister supports flexible-grained paper search via hierarchical register, while traditional method fails due to abstract cannot contain required details.
}
\label{fig:head}
\end{figure}

%% file: figs/method.tex
\begin{figure*}[t!]
\centering
\includegraphics[width=\linewidth]{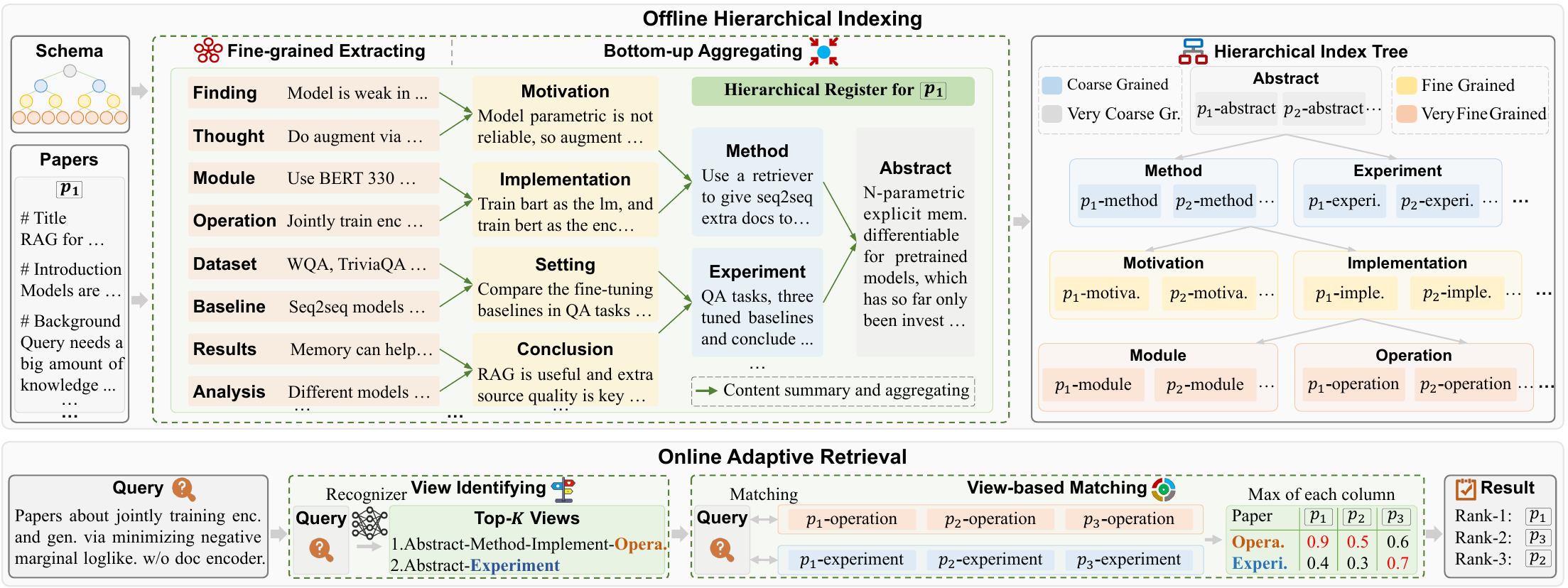} 
\caption{PaperRegister includes hierarchical indexing and  adaptive retrieval. Offline, PaperRegister constructs hierarchical index tree via fine-grained content extracting and bottom-up content aggregating based on  a hierarchical register schema. Online, PaperRegister first identify views of query and then conduct view-based matching.}  
\label{fig:method}
\end{figure*}

%% file: sections/register.tex
\section{PaperRegister}

Existing paper search systems fail to handle queries at flexible granularity due to they primarily collect paper abstract to construct index of corpus, which does not contain detailed information to support queries at finer granularity. 
To this end, we propose PaperRegister, which can support  flexible-grained paper search via hierarchical register indexing. 
As illustrated in Figure~\ref{fig:method}, 
in the offline stage, a hierarchical index tree is built for paper corpus, and in the online stage, adaptive retrieval is performed by selecting and using suitable indexes from the tree.

\subsection{Task Formulation}
The task involved in this work provides a query $q$ as input, with the goal of retrieving $M$ relevant papers $\{p^{(m)}\}_{m=1}^M$ based on the index $\mathcal{I}$ of  paper corpus $\mathcal{C}$, which can be expressed as  follows:
\begin{equation}
    \{p^{(m)}\}_{m=1}^M = \mathcal{F}(q,\mathcal{I})
\end{equation} 

For regular paper search, view of query $q$ is typically a coarse-grained topic that user is interested in, and systems mainly construct index $\mathcal{I}$ based on paper abstract to match against $q$. 
As user's requirements become more flexible, the view may refer to details such as module configuration and training operation. 
In this context, since this detailed information typically do not appear in abstract, existing paper search systems cannot retrieve accurately.

\subsection{Offline Hierarchical Indexing}
Considering the granularity of query view can be flexible across very fine-grained, moderately fine-grained, to coarse-grained, 
PaperRegister offline constructs a hierarchical index tree for paper corpus to support flexible paper search. 
Specifically, based on a hierarchical register schema, PaperRegister obtains hierarchical register for each paper through  fine-grained content extracting and  bottom-up content aggregating, and then hierarchical index tree is constructed by merging register of all papers. 

\paragraph{Hierarchical Register Schema.}
To obtain the hierarchical register for each paper, we first design a hierarchical register schema, which consists of information nodes as the following formula:
\begin{equation}
    N_{i,j} = \{n_{i,j}: (c_{i,j},~\{N_{i+1,j^{\prime}}\}_{j^{\prime}=1}^Z)\}
\end{equation} where $N_{i,j}$ is the $j$-th information node at the $i$-th layer, $n_{i,j}$ is node name, $c_{i,j}$ is expected node content from specific paper, $N_{i+1,j^{\prime}}$ is a sub-node of $N_{i,j}$, $Z$ is number of sub-nodes under each node, and $i\in[1,...,L]$,  $L$ is number of layers in schema.

In hierarchical register schema,  shallow-layer nodes represent coarse-grained information, and deep-layer nodes represent finer-grained information.
For example, in the $i\!+\!1$-th layer, $N_{i+1,j\prime}$ and $N_{i+1,j\prime+1}$  respectively denote \textit{module} configuration and  training \textit{operation} of paper, which are two kinds of relatively fine-grained information. While their common parent in the $i$-th layer, $N_{i,j}$, denotes method \textit{implementation}, which is a  coarser-grained summary of $N_{i+1,j\prime}$ and $N_{i+1,j\prime+1}$, thereby forming schema with hierarchical-dependency nodes. 

In addition, since different types of papers have different styles and formats, we design five kinds of schema, corresponding to five paper types, and  use a large language model to determine the type of each paper. Details are presented in Appendix~\ref{app_a}.

\paragraph{Fine-grained Content Extracting.} 
Based on the above hierarchical register schema, in order to obtain fine-grained contents as much as possible, PaperRegister first extracts content for each information node at the $L$-th layer from paper $p^{(m)}$, which can be represented as the following formula:
\begin{equation}
    c_{L,j}^{(m)}=\mathcal{M}_{extract}(p^{(m)}, n_{L,j})
\end{equation}
where $n_{L,j}$ represents name of information node $N_{L,j}$ and $\mathcal{M}_{extract}$ is the extracting module.

To achieve accurate extraction, learning from several widely recognized works, in which large language models are proven to possess reliable capabilities for content extraction tasks~\cite{edge2024local,li2025deepsolution}, PaperRegister employs a large language model as extracting module. 
Specifically, PaperRegister takes the node name and text-formatted paper as input, uses instructions to guide in outputting corresponding content for the information node, and leaves it blank if the paper does not include corresponding content.
The detailed process and instructions are in the Appendix~\ref{app_b}.

\input{figs/train}

\paragraph{Bottom-up Content Aggregating.}
After extracting all the finest-grained contents, in order to smoothly obtain a complete hierarchical register, PaperRegister aggregates node contents layer by layer from bottom to top based on the hierarchical register schema,  represented as following formula:
\begin{equation}
    c_{i,j}^{(m)}=\mathcal{M}_{aggregate}(\{c_{i+1,j \prime}^{(m)}\}_{j\prime=1}^{Z})
\end{equation} where the layer $i$ is from $L-1$ to 1, thereby obtaining content for each information node in the hierarchical register schema for the  paper $p^{(m)}$.

To achieve accurate aggregation, like extraction process, PaperRegister uses a large language model as the aggregating module $\mathcal{M}_{aggregate}$.
Specifically, PaperRegister takes contents of sub-nodes as input and uses instructions to guide large language model in summarizing, condensing, and removing details to turn into summary text, thus obtaining content for upper-layer information node.
The detailed process and instructions are in Appendix~\ref{app_c}.

At this point, PaperRegister obtains node contents at various granular levels to compose the hierarchical register for $p^{(m)}$. And then merges register of all papers in corpus $\mathcal{C}$ to construct  hierarchical index tree $\mathcal{I}_{h}$, as shown in following formula:
\setlength\abovedisplayskip{5pt}
\begin{equation}
\resizebox{.8\hsize}{!}{$\mathcal{I}_{h}=\{\{ {\rm  I}_{i,j}\}_{j=1}^Z\}_{i=1}^L=\{\{\mathcal{M}_{idx}\{c_{i,j}^{(m)}\}_{m=1}^{|\mathcal{C}|}\}_{j=1}^{Z}\}_{i=1}^{L}$}
\end{equation} 
where $\mathcal{M}_{idx}$ is the indexing module such as BM25 and DPR,  each ${\rm  I}_{i,j}$ is an index in hierarchical index tree, corresponding to a kind of view for corpus.

\subsection{Online Adaptive Retrieval}

Based on the offline hierarchical indexing above, PaperRegister perform online adaptive retrieval via suitable indexes from hierarchical index tree.
Specifically, PaperRegister first identifies  views involved in the  query, determining corresponding indexes from hierarchical index tree, then conduct retrieval by matching query against these indexes.

\paragraph{View Identifying.} 
To achieve adaptive retrieval, PaperRegister first uses a view recognizer to identify the view $v_k$ in query $q$, represented as follows:
\begin{equation}
    \{{v}_{k}\}_{k=1}^K=\mathcal{M}_{identify}(q)
\end{equation} where the candidate set of ${v}_k$ is all node paths in the hierarchical register schema, $\mathcal{M}_{identify}$ is employed base on a small-scale language model with special training, which will be explained in next section. 
And to ensure that identifying results can cover the real views of query as more as possible, $\mathcal{M}_{identify}$ uses the beam search strategy~\cite{Holtzman2020The} to sample the top-$K$ output views.

\paragraph{View-based Matching.}
After identifying views in query, PaperRegister looks up corresponding indexes from hierarchical index tree $\mathcal{I}_{h}$, as follows:
\begin{equation}
\{{\rm I}_k\}_{k=1}^K = \mathcal{M}_{lookup}(\mathcal{I}_{h},\{{v}_{k}\}_{k=1}^K)
\label{for:v}
\end{equation} where ${\rm I}_k$ is consisted of   $\{c_{k}^{(m)}\}_{m=1}^{|\mathcal{C}|}$ as Formula 5. 

Then these indexes are used to calculate relevance score of input query $q$ and each paper $p^{(m)}$~~:
\begin{equation}
    s(q,p^{(m)}) = \max\{\mathcal{M}_{rel}(q,~{c}_{k}^{(m)})_{k=1}^K\}
\label{for:S}
\end{equation} where $\mathcal{M}_{rel}$ is   relevance module such as BM25. 

Finally, papers with top-$M$ relevance score, $\{p^{(m)}\}_{m=1}^M$, are selected as final results for  query.

%% file: figs/train.tex
\begin{figure*}[t!]
\centering
\includegraphics[width=\linewidth]{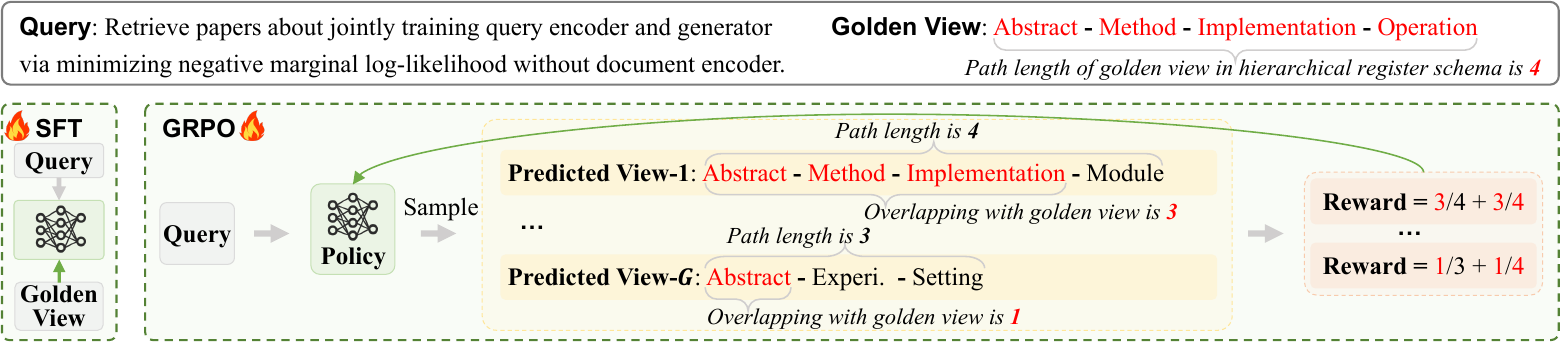} 
\caption{Illustration of view recognizer training,   including SFT and GRPO via hierarchical reward, which is calculated based on the closeness level  of predicted view and golden view in the hierarchical register schema.}
\label{fig:train}
\end{figure*}

%% file: sections/recognizer.tex
\section{View Recognizer Training}
For PaperRegister system, view recognizer is a key module, which must both alleviate latency and ensure accuracy.
To this end, as in Figure~\ref{fig:train}, we use a small-scale language model with 0.6B parameters as base, first do  supervised fine-tuning (SFT), and then further enhance capability by hierarchical-reward group relative policy optimization (GRPO).

\paragraph{Training Data.} 
The format of training dataset to support SFT and GRPO is shown as follows:
\begin{equation}
     \mathcal{D}_{train} = \{q_{j}, v_{j}\}_{j=1}^{|\mathcal{D}_{train}|}
\label{for:train_dataset}
\end{equation} where $q_{j}$ is a query, $v_{j}$ is golden view of query, which is node path in hierarchical register schema. 

\paragraph{Supervised Fine-tuning.}
To give small-scale view recognizer basic identifying ability and make it easier for subsequent reinforcement learning, we first perform SFT  by minimizing following loss:
\setlength\abovedisplayskip{5pt}
\begin{equation}
\resizebox{.8\hsize}{!}{$\mathcal{L}_{SFT}(\theta)=-\mathbb{E}_{(q,v) \sim \mathcal{D}_{train}} \sum_{t=1}^{|v|} \log \pi_{\theta}(v_t|q, v_{<t})$}\end{equation}
where $\pi_{\theta}$ is model,  $v_t$ is the $t$-th golden view token.

\paragraph{Hierarchical-reward GRPO.}
To strengthen the capability of small-scale view recognizer, we  conduct GRPO~\cite{Shao2024DeepSeekMathPT} with hierarchical reward, which is calculated via the  closeness level  of  predicted view and golden view in hierarchical register schema. 
Specially, considering the hierarchical-dependency feature of  register schema, different wrong predicted views should not be treated equally. For example, the golden view of query is represented as \textit{Abstract-Method-Implementation-Operation} in schema,  and two predicted views are \textit{Abstract-Method-Implementation-Module} and \textit{Abstract-Experiment-Dataset} respectively. Although both are wrong, the first one is better than the second because it is closer to golden view in hierarchical register schema, suggesting a higher information overlap with golden view.

Based on this, we measure path overlap between  predicted view $\hat{v}_{j}$ and golden view $v_{j}$ in  hierarchical register schema, obtaining hierarchical reward:
\begin{equation}
    r = \frac{\mathcal{M}_{overlap}(v_{j},\hat{v}_{j})}{|\hat{v}_{j}|} + \frac{\mathcal{M}_{overlap}(v_{j},\hat{v}_{j})}{|v_{j}|}
\end{equation} where $\mathcal{M}_{overlap}$ returns  overlap level of inputs, $|\hat{v}_{j}|$ means  path length of predicted view in hierarchical register schema, and $|v_{j}|$ is that of golden view.

Then this hierarchical reward  is used in GRPO, which is by maximizing the following objective:
\setlength\abovedisplayskip{5pt}
\begin{equation}
\resizebox{.8\hsize}{!}{$\begin{aligned}
\mathcal{J}_{GRPO}(\theta) ={} & \mathbb{E}_{\left[q \sim D_{train}, \left\{v_{i}\right\}_{i=1}^G \sim \pi_{\theta_{\text{old}}}(v \mid q)\right]}  \frac{1}{G} \sum_{i=1}^G \frac{1}  {\left|v_{i}\right|}  \sum_{t=1}^{\left|v_{i}\right|} \Biggl\{ \\
& \min  \Biggl[ 
\frac{\pi_\theta\left(v_{i,t} \mid q, v_{i,<t}\right)}{\pi_{\theta_{\text{old}}}\left(v_{i,t} \mid q, v_{i,<t}\right)} \hat{A}_{i,t}, \text{clip} ( \frac{\pi_\theta\left(v_{i,t} \mid q, v_{i,<t}\right)}{\pi_{\theta_{\text{old}}}\left(v_{i,t} \mid q, v_{i,<t}\right)}, 1 - \epsilon, 1 + \epsilon ) \hat{A}_{i,t} \Biggr] \\
&  - \beta \mathbb{D}_{\text{KL}} \left[ \pi_\theta \parallel \pi_{\text{ref}} \right] \Biggr\}, ~\text{where}~\hat{A}_{i,t}=\frac{r_{i} - mean(\{r_{1},r_{2},...,r_{G}\})}{std(\{r_{1},r_{2},...,r_{G}\})}
\end{aligned}$}
\end{equation}
where $\pi_{\theta_{\text{old}}}$ is initialized by SFT model, $G$ is  number of each group, $\epsilon$ and $\beta$ are hyper-parameters.


%% file: sections/experiment.tex
\input{tabs/main}

\section{Experiments}

\subsection{Experimental Settings}
\paragraph{Datasets.}
To conduct experiments for flexible-grained  paper search, 
we first use  \textit{LitSearch}~\cite{ajith2024litsearch},  which mainly contains coarse-grained queries, and then we build \textit{Flexible-grained Search}, a new dataset covering queries across various granularity, including data for test, development and training. 
The metrics reported in experiments are \textbf{R@5} (recall@5) and \textbf{R@10} (recall@10). 
And detailed statistical information is in the Appendix~\ref{app_d}. 

(1) \textbf{LitSearch}~\cite{ajith2024litsearch}. By using GPT-4 to rewrite citations in papers and asking authors to write, this dataset constructs 597 paper search queries, mainly involving coarse-grained topics like \textit{``Where can I find some researches on evaluating consistency in generated summaries''}.

(2) \textbf{Flexible-grained Search}. We  collect 4,200 papers in the arXiv platform as corpus and perform the offline hierarchical indexing.
To build training and development data, we pick 2,100 paper registers, and generate query based on each node content in each register by LLM, where the node path is  golden view of query. 
Then we randomly split, getting 13,824 training data and 695 development data, training data are used in view recognizer training, and development data are for analysis in Table~\ref{table:router}. The format of training and development data is $\{q_j, v_j, p_j\}$.
To build test data across various granularity, we pick the other 2,100 paper registers. And in order to prevent data leakage, we do not directly use node content to generate query. Instead, we employ LLM to find original paper text related to each node, then generate queries based on these text, obtaining 5,644 test data. Thanks to the hierarchical feature of  register, test data cover: \textbf{F.g.Search-1} (general granularity), \textbf{F.g.Search-2} (fine granularity), and \textbf{F.g.Search-3} (very fine granularity). 
Format of each test data is $\{q_j, p_j\}$,
and LLM in this section is Qwen3-32B~\cite{qwen3}.

\paragraph{Selected Baselines.} We select baselines from commonly used or advanced methods suitable for paper search tasks.
(1) \textbf{Direct Matching.} These methods directly use the paper title or abstract to construct corpus index, and then perform retrieval by matching query with these contents. 
In addition, to demonstrate that the effectiveness of PaperRegister does not simply stem from using total paper text, we include a method that constructs the corpus index using the total paper text into this category of baselines.
(2) \textbf{Query Enhancing.} These methods aim to improve paper search performance by paraphrasing the input query, including: Rewriting~\cite{ma2023query}, which uses a LLM to rewrite original query. HyDE~\cite{gao2023precise}, which uses a LLM to generate a fake document based on input query and retrieves real documents by this fake document. CSQE~\cite{lei2024corpus}, which initially retrieves several documents and then uses a LLM to expand original query based on these initially-retrieved documents.
We use Qwen3-32B as LLM for these baselines.
(3) \textbf{Multi-field Indexing.} These methods split original document into multiple parts, get flat multi-field index, then calculate similarity by selected indexes and integrate to determine overall relevance~\cite{li2025multifield}. In experiments, 
due to the lack of a universal multiple-filed partitioning for all kinds of paper,
we employ four settings, including splitting the original paper to fixed length of 512-token chunk or by raw paragraph, taking average similarity or maximum similarity of all parts as final score,  represented as $\text{Chunk}_{avg}$, $\text{Chunk}_{max}$, $\text{Paragraph}_{avg}$, and $\text{Paragraph}_{max}$, respectively.

\paragraph{Implementation Details.} 
In the offline stage, we employ Qwen3-32B~\cite{qwen3} as the large language model in process of fine-grained content extracting and bottom-up content aggregating, and deploy it as API by vllm\footnote{{https://pypi.org/project/vllm/}} on two A-100 80G GPU for convenience. 
In the online stage, we set $K$ as 5 and $M$ as 5 or 10, use Qwen3-0.6B~\cite{qwen3} as the base model of view recognizer, employ  a prefix tree-based restricted decoding strategy~\cite{tang2024selfretrieval} in view identifying process to prevent the model outputting irrelevant token, and conduct view-based matching process via rank-bm25\footnote{{https://pypi.org/project/rank-bm25/}} for BM25 and gte-Qwen2-7B-instruct~\cite{li2023towards}  for DPR.
For the view recognizer training, we use TRL\footnote{https://github.com/huggingface/trl} framework to conduct SFT and GRPO, with train epoch as 5 in SFT, $G$ as 5 and train epoch as 2 in GRPO, and all other parameters as default value. 

\input{tabs/ablation}

\subsection{Overall Results} 
Results compared with baselines are shown in the Table~\ref{tab:main_results}, there are two main conclusions:

(1) \textbf{PaperRegister is an effective system to addressing flexible-grained paper search.} 
As shown in the Table~\ref{tab:main_results}, PaperRegister demonstrates excellent performance in paper search tasks at various granularity, outperforming all baselines in both BM25-based matching and DPR-based matching settings, on both recall@5 and recall@10 metrics. 
For example, in F.g.Search-3, under the DPR-based matching setting, PaperRegister achieves a recall@5 score of 80.8, while using abstract-based index yields only 58.2, where PaperRegister can achieve a performance improvement of 22.6. 
All in all, PaperRegister can improve a lot compared with various previous paper search methods.

(2) \textbf{Compared to traditional methods, advantage of PaperRegister become more pronounced at finer granular queries.} 
According to experimental setting, from LitSearch to F.g.Search-1 to F.g.Search-2 to F.g.Search-3, the granularity of query becomes increasingly finer. Experimental results show performance improvement of PaperRegister over abstract-based method becomes larger.
For example, under the DPR-based matching setting, PaperRegister achieves recall@5 scores of 81.0, 84.1, 79.9, and 80.8 on the four datasets, respectively, while using the abstract-based index yields scores of 77.7, 69.4, 62.1, and 58.2. Here, PaperRegister delivers performance improvements of 3.3, 14.7, 17.8, and 22.6, respectively.
This can to some extent validate rationality of our motivation, traditional  paper search methods via abstract-based index struggles to handle fine-grained queries.

\subsection{Ablation Results}
To validate importance of hierarchical register indexing, we employ ablation by simplifying register schema. 
As shown in Table~\ref{tab:ablation}, where ``w/ only layer-1'' means retaining only the coarsest-grained nodes, ``w/ only layer-2'' means retaining only the medium-grained nodes, and ``w/ only layer-3'' means retaining only the finest-grained nodes.
Base on the ablation results, we get the following conclusions:

(1) \textbf{Hierarchical register indexing is important for ensuring accurate paper search.}
Compared to using complete hierarchical register schema, using a schema composed of nodes from any single layer leads to obvious performance drop. For example, recall@5 of LitSearch under BM25-based matching drops from 69.7 to 64.5, 66.8, 64.6 for ``w/ only layer-1'', ``w/ only layer-2'', ``w/ only layer-3'', respectively. And other datasets also show similar situation.
Therefore, results strongly prove the importance of hierarchical register indexing.

\input{figs/router}

(2) \textbf{Different layers in hierarchical register schema serve queries at different granularity.}
For example, under ``w/ only layer-1'' setting, the DPR-based matching and recall@5 metric of F.g.Search-1 drops from 84.1 to 82.5 (1.6 decrease), while F.g.Search-1 is from 80.8 to 68.1 (12.7 decrease), which is significantly larger than 1.6. Conversely, under ``w/ only layer-3'' setting, F.g.Search-1 drops from 84.1 to 77.8 (6.3 decrease), while F.g.Search-3 is from 80.8 to 80.4 (0.4 decrease), which is much smaller than 6.3.
Therefore, nodes of different layers play specific roles in handling queries of different granularity, further strongly proving necessity of hierarchical register indexing.

\subsection{Detailed Analysis}
In this section, we conduct detailed analysis to explore the compatibility and real-world practicality of PaperRegister. Due to the space constraint, we only report DPR-based performance for analysis.

\paragraph{Effect of View Recognizer and Training.}
In order to strictly analyze the role of view recognizer in PaperRegister and the necessity of our training, we first examine relationship between the capability of view recognizer and  the final performance of PaperRegister in Figure~\ref{fig:router}, then compare with Qwen3-32B and conduct ablation in Table~\ref{table:router}. The detailed  process and findings are in following:

As shown in Figure~\ref{fig:router}, where ``Bad Recognizer'',``Random Recognizer'',``Weak Recognizer'' represent the completely incorrect, randomly predicting, and weak-performing view recognizers, respectively, the curve demonstrates a clear positive correlation between the capability of view recognizer and final performance of PaperRegister. 
\textit{Therefore, building an accurate view recognizer is a key factor for the PaperRegister system to achieve high performance in complex paper search.}

\input{tabs/router}

As shown in Table~\ref{table:router}, 
We compare accuracy and latency of view identifying between the view recognizer in PaperRegister and Qwen3-32B (enable thinking) under zero-shot and few-shot settings. The results show the view recognizer in PaperRegister is with better accuracy and latency than powerful Qwen3-32B.
Furthermore, we only keep SFT for the view recognizer training, or replace the hierarchical reward with direct 0-1 reward in GRPO training. Results show that accuracy decreases to some extent.
\textit{Therefore, hierarchical-reward GRPO is effective to enhance the view recognizer achieving both high performance and low latency.}

\paragraph{Compatibility with PaSa.}
Given that some complex information retrieval frameworks such as PaSa~\cite{he2025pasa} incorporate various modules like rewriting, retrieval, iteration, and filtering, to explore whether PaperRegister can be compatible with such complex frameworks and further enhance  paper search performance, we conduct experiment by replacing the original search module in PaSa with PaperRegister.
As shown in Figure~\ref{fig:pasa}, PaperRegister can further improve  performance of PaSa on paper search tasks across various granularity. \textit{Therefore, PaperRegister can be effectively adapted as a search system into complex frameworks and further enhance the paper search capability.}

\input{figs/pasa}

\paragraph{Online Search Efficiency.} 
Considering that the efficiency of  online paper search system is a very crucial aspect for ensuring its usability in real-world scenarios, we compare the online search efficiency of PaperRegister with multiple baselines.
As shown in the Table~\ref{table:effiency}, the online latency of these baselines is significantly higher than that of PaperRegister, which limits the practical applicability of these methods in real-world scenarios. \textit{In contrast, PaperRegister demonstrates relatively acceptable online latency, proving it has a good potential for application in the real-world paper search tasks.}

\paragraph{Reducing Indexing Time.} 
Although PaperRegister achieves excellent performance, it requires a relatively long time for offline indexing. Therefore, we explore whether it can sacrifice a certain amount of performance to reduce time consumption. As shown in Table~\ref{table:tradeoff}, we replace original two-step registering (extracting and aggregating) with directly extracting all contents. Results show this can significantly reduce the offline indexing time with acceptable performance degradation, which \textit{provides an alternative approach  allowing a trade-off based on specific requirements in real-world application}.

%% file: tabs/main.tex
\begin{table*}[t!]
\centering
\resizebox{\linewidth}{!}{
\begin{tabular}{lcccccccccccccccc}
\toprule

\multirow{4}[1]{*}{\textbf{Method}} & \multicolumn{8}{c}{\textbf{BM25-based Paper Search}} & \multicolumn{8}{c}{\textbf{DPR-based Paper Search}} \\

\cmidrule(lr){2-9} \cmidrule(lr){10-17}

& \multicolumn{2}{c}{\textbf{LitSearch}} & \multicolumn{2}{c}{\textbf{F.g.Search-1}} & \multicolumn{2}{c}{\textbf{F.g.Search-2}} & \multicolumn{2}{c}{\textbf{F.g.Search-3}} & \multicolumn{2}{c}{\textbf{LitSearch}} & \multicolumn{2}{c}{\textbf{F.g.Search-1}} & \multicolumn{2}{c}{\textbf{F.g.Search-2}} & \multicolumn{2}{c}{\textbf{F.g.Search-3}} \\

\cmidrule(lr){2-3} \cmidrule(lr){4-5}  \cmidrule(lr){6-7} \cmidrule(lr){8-9} \cmidrule(lr){10-11} \cmidrule(lr){12-13}  \cmidrule(lr){14-15} \cmidrule(lr){16-17}
& R@5 & R@10 & R@5 & R@10 & R@5 & R@10 & R@5 & R@10 & R@5 & R@10 & R@5 & R@10 & R@5 & R@10 & R@5 & R@10 \\

\midrule
\rowcolor[rgb]{ .906,  .902,  .902} \multicolumn{17}{c}{Direct Matching} \\
\midrule
Title                                              & 54.2          & 59.5          & 42.0          & 47.3          & 36.1          & 42.0          & 30.4          & 35.3          & 66.8          & 73.1          & 52.0          & 58.8          & 45.8          & 52.1          & 40.6          & 47.1          \\
Abstract                                           & 67.1          & 71.3          & 63.7          & 69.0          & 57.4          & 62.9          & 54.2          & 59.7          & 77.7          & 83.2          & 69.4          & 74.3          & 62.1          & 68.1          & 58.2          & 63.1          \\
Total Paper                                        & 64.2          & 68.9          & 79.7          & 82.4          & 81.4          & 83.5          & 84.2          & 86.2          & 74.8          & 81.9          & 72.9          & 78.7          & 68.8          & 73.8          & 65.8          & 71.0          \\
\midrule
\rowcolor[rgb]{ .906,  .902,  .902} \multicolumn{17}{c}{Query Enhancing} \\
\midrule
Rewriting~\cite{ma2023query}
& 61.9          & 69.6          & 58.4          & 64.2          & 54.1          & 59.8          & 50.9          & 56.3          & 76.0          & 83.2          & 67.1          & 72.6          & 60.4          & 65.2          & 55.2          & 60.9          \\
HyDE~\cite{gao2023precise}
& 68.9          & 75.4          & 60.3          & 66.3          & 54.3          & 60.0          & 51.5          & 57.6          & 76.5          & 83.4          & 66.8          & 72.3          & 58.9          & 65.7          & 56.3          & 62.4          \\
CSQE~\cite{lei2024corpus}
& 69.0          & 73.8          & 59.1          & 64.4          & 54.2          & 58.6          & 51.3          & 55.9          & 77.9          & 81.7          & 65.8          & 71.1          & 60.2          & 65.5          & 55.2          & 60.9          \\

\midrule
\rowcolor[rgb]{ .906,  .902,  .902} \multicolumn{17}{c}{Multi-field Indexing} \\
\midrule

$\text{Chunk}_{avg}$~\cite{li2025multifield}                               & 58.1          & 67.6          & 68.1          & 74.2          & 66.6          & 71.9          & 65.9          & 71.6          & 49.7          & 58.6          & 51.1          & 58.3          & 46.4          & 53.4          & 46.4          & 52.2          \\
$\text{Chunk}_{max}$~\cite{li2025multifield}                               & 67.9          & 75.5          & 80.0          & 83.1          & 81.4          & 84.7          & 85.3          & 87.6          & 72.6          & 79.7          & 79.8          & 83.1          & 76.6          & 80.5          & 79.0          & 82.1          \\
$\text{Paragraph}_{avg}$~\cite{li2025multifield}                           & 29.7          & 38.3          & 58.9          & 66.6          & 59.8          & 66.5          & 58.2          & 66.2          & 20.5          & 22.8          & 23.3          & 30.6          & 23.1          & 29.8          & 22.4          & 29.2          \\
$\text{Paragraph}_{max}$~\cite{li2025multifield}                           & 64.3          & 70.9          & 73.8          & 78.8          & 75.4          & 79.9          & 80.8          & 83.9          & 79.5          & 85.0          & 79.2          & 82.8          & 76.5          & 81.4          & 78.8          & 81.8          \\

\midrule
\rowcolor[rgb]{ .906,  .902,  .902} \multicolumn{17}{c}{Hierarchical Register Indexing} \\
\midrule

PaperRegister (Ours)                               & \textbf{69.7} & \textbf{76.4} & \textbf{89.7}* & \textbf{90.9}* & \textbf{88.0}* & \textbf{89.0}* & \textbf{87.5}* & \textbf{88.7}* & \textbf{81.0}* & \textbf{87.1}* & \textbf{84.1}* & \textbf{87.1}* & \textbf{79.9}* & \textbf{82.5}* & \textbf{80.8}* & \textbf{82.9}* \\

\bottomrule
\end{tabular}%
}
\caption{Main results on paper search across various granularity, where granularity is coarse-to-fine from  LitSearch to F.g.Search-3. Results shows that PaperRegister is an effective system for flexible-grained paper search and the advantage of PaperRegister become more pronounced at finer granular tasks. The * indicates statistical significance (p < 0.05) compared with the best baseline. And experiment results by more metrics are reported in Appendix~\ref{app_e}.}
\label{tab:main_results}%
\end{table*}%

%% file: tabs/ablation.tex
\begin{table*}[t!]
\centering
\resizebox{\linewidth}{!}{
\begin{tabular}{lcccccccccccccccc}
\toprule

\multirow{4}[1]{*}{\textbf{Method}} & \multicolumn{8}{c}{\textbf{BM25-based Paper Search}} & \multicolumn{8}{c}{\textbf{DPR-based Paper Search}} \\

\cmidrule(lr){2-9} \cmidrule(lr){10-17}

& \multicolumn{2}{c}{\textbf{LitSearch}} & \multicolumn{2}{c}{\textbf{F.g.Search-1}} & \multicolumn{2}{c}{\textbf{F.g.Search-2}} & \multicolumn{2}{c}{\textbf{F.g.Search-3}} & \multicolumn{2}{c}{\textbf{LitSearch}} & \multicolumn{2}{c}{\textbf{F.g.Search-1}} & \multicolumn{2}{c}{\textbf{F.g.Search-2}} & \multicolumn{2}{c}{\textbf{F.g.Search-3}} \\

\cmidrule(lr){2-3} \cmidrule(lr){4-5}  \cmidrule(lr){6-7} \cmidrule(lr){8-9} \cmidrule(lr){10-11} \cmidrule(lr){12-13}  \cmidrule(lr){14-15} \cmidrule(lr){16-17}
& R@5 & R@10 & R@5 & R@10 & R@5 & R@10 & R@5 & R@10 & R@5 & R@10 & R@5 & R@10 & R@5 & R@10 & R@5 & R@10 \\

\midrule

PaperRegister & \textbf{69.7} & \textbf{76.4} & \textbf{89.7} & \textbf{90.9} & \textbf{88.0} & \textbf{89.0} & \textbf{87.5} & \textbf{88.7} & \textbf{81.0} & \textbf{87.1} & \textbf{84.1} & \textbf{87.1} & \textbf{79.9} & \textbf{82.5} & \textbf{80.8} & \textbf{82.9} \\
    
~~w/ only layer-1 & 64.5 & 70.4 & 87.8 & 90.6 & 77.4 & 80.2 & 73.5 & 76.7 & 79.0 & 83.5 & 82.5 & 85.2 & 73.4 & 77.1 & 68.1 & 72.2 \\
~~w/ only layer-2 & 66.8 & 73.8 & 84.4 & 86.5 & 87.1 & 88.8 & 82.0 & 84.4 & 78.9 & 84.2 & 81.1 & 84.9 & 79.3 & 82.2 & 73.5 & 77.9 \\
~~w/ only layer-3 & 64.6 & 71.7 & 79.9 & 81.8 & 83.7 & 85.0 & 86.8 & 88.1 & 78.8 & 85.2 & 77.8 & 81.1 & 77.3 & 80.0 & 80.4 & 82.6 \\

\bottomrule
\end{tabular}%
}
\caption{Ablation for hierarchical register indexing. Results show that hierarchical index tree is important for flexible-grained paper search and different layers in  hierarchical register schema serve queries at different granularity.}
\label{tab:ablation}%
\end{table*}%

%% file: figs/router.tex
\begin{figure}[t!]
\centering
\includegraphics[width=\linewidth]{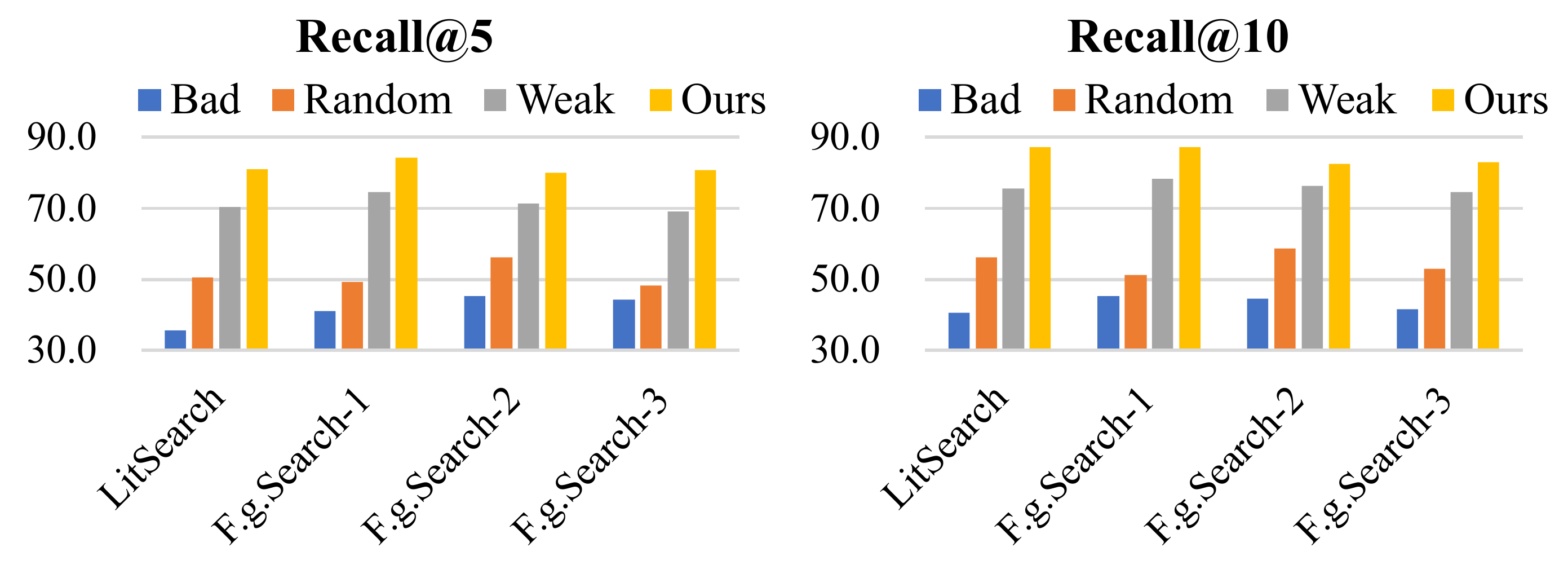} 
\caption{Performance of PaperRegister with different view recognizer. The figure shows a strong recognizer is with obvious positive impact on the overall system.} 
\label{fig:router}
\end{figure}


%% file: tabs/router.tex
\begin{table}[t]
\centering  
\tiny
\resizebox{0.99\linewidth}{!}{
\begin{tabular}{lcc}
\toprule
\textbf{Recognizer} & \textbf{ACC} $\uparrow$ & \textbf{Latency} (s) $\downarrow$ \\
\midrule
Qwen3-32B  & 30.5 & 28.3  \\ 
Qwen3-32B (few-shot) & 47.8 & 37.8  \\ 
View Recognizer in PaperRegister & \textbf{83.5} & \textbf{2.3} \\ 
~~w/~~ only SFT & 80.9 & -  \\ 
~~w/o hierarchical reward & 81.7  & -  \\ 
\bottomrule
\end{tabular}
}
\caption{Comparison and ablation for the view recognizer training. The table shows that the training process in this work is an effective approach to obtain a view recognizer with both high performance and low latency.}
\label{table:router}
\end{table}


%% file: figs/pasa.tex
\begin{figure}[t!]
\centering
\includegraphics[width=0.95\linewidth]{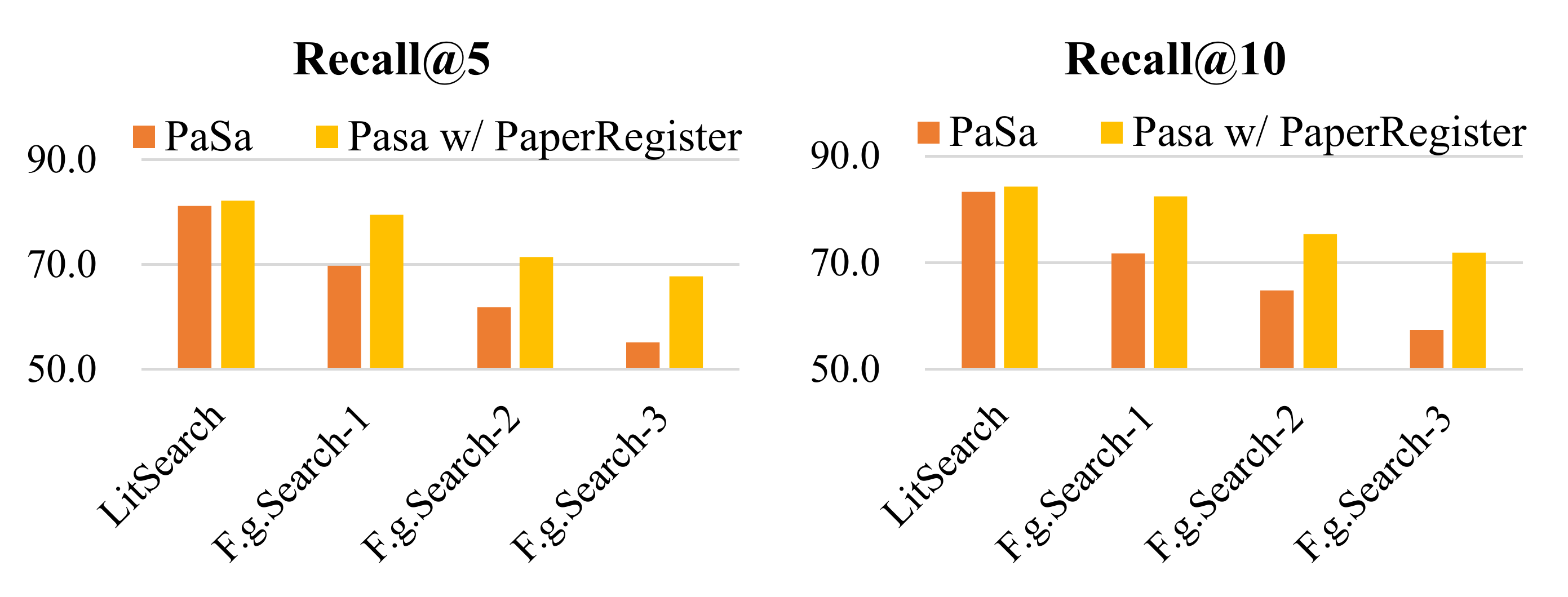} 
\caption{Performance of combining PaperRegister into PaSa framework. The figure shows that PaperRegister can greatly cooperate with complex modules in PaSa.} 
\label{fig:pasa}
\end{figure}

%% file: sections/related.tex
\section{Related Work}

\paragraph{Query Enhancing.} Previous works mainly improve paper search by query enhancing, such as using LLM rewriting~\cite{ma2023query,anand2023context}, generating pseudo-documents to replace original query~\cite{gao2023precise,li2024automir}, employing powerful models as agent system for multi-round  expansion~\cite{he2025pasa,ren2025towards}, augmenting via initially retrieved documents~\cite{lei2024corpus,li2025pseudo}, or extracting keywords from corpus to enrich  query~\cite{kang2024taxonomy,zhang2025scientific}. 
Although these methods can improve regular paper search, they fail to address flexible-grained paper search because do not solve  fundamental flaw of abstract-based index.

\paragraph{Multi-filed Indexing.} Recently, several studies improve by splitting documents into multiple parts to build a flat, multi-field index~\cite{takeshita-etal-2024-gengo,li2025multifield,shi2025hypercube,chen2025enrichindex}. However, these works are mainly limited to a single granularity level, still struggling to adapt to paper search that demands varying degrees of granularity.
In contrast, PaperRegister fundamentally differs by introducing a hierarchical register index. Tree-structured organization contains information across various granularity, enabling system to conduct matching at the appropriate levels, thereby effectively supporting flexible-grained paper search.


%% file: sections/conclusion.tex
\input{tabs/effiency}

\input{tabs/tradeoff}

\section{Conclusion}
In this work, we propose PaperRegister, which can support paper search queries at flexible granularity. 
Specially, PaperRegister offline constructs a hierarchical index tree and online adaptively retrieve based on this tree.
Furthermore, we design a powerful and low-latency view recognizer by applying supervised finetuning and hierarchical-reward group relative policy optimization.
Experiments on flexible-grained paper search tasks demonstrate PaperRegister is an effective solution, with improvement being more pronounced as granularity finer.
This work offers a promising direction for developing more powerful paper search systems in future.

\section{Limitations}
PaperRegister transforms raw paper into hierarchical paper register, thereby achieving strong performance in flexible-grained paper search. 
The main limitations are on the consumption associated with LLM utilization and risks stemming from the inherent limitations of LLM.
Firstly, using LLM for offline hierarchical indexing requires substantial computational and storage resources. When the paper corpus is large, this may place high demands on computer hardware.
Secondly, our experiments are conducted only on normal paper corpus, where LLM could effectively accomplish the specified task. We have not yet do validate under extreme conditions, such as with incomplete papers or papers from niche domains, primarily due to a lack of relevant experimental data.
Therefore, mitigating computational resource consumption of PaperRegister, as well as verifying and improving effectiveness and robustness across a broader range of paper, remain important directions for the future work.

%% file: tabs/effiency.tex

\begin{table}[t]
\centering  
\tiny
\begin{tabularx}{\linewidth}{Xcccc}
\toprule
\textbf{Method} & \textbf{Avg. R@5} & \textbf{Avg. R@10} & \textbf{Online Latency } (s) $\downarrow$  \\
\midrule
Rewriting & 64.7 & 70.5 & 9.3    \\ 
HyDE & 64.6 & 70.9  & 20.7   \\ 
CSQE  & 64.8 & 69.8  & 33.5  \\ 
$\text{Chunk}_{max}$ & 77.0 & 81.4  & 5.4  \\ 
PaperRegister (Ours) & \textbf{81.5} & \textbf{84.9}  & \textbf{2.5}   \\ 
\bottomrule
\end{tabularx}
\caption{Comparison of online latency. PaperRegister is with less latency for better real-world applicability.}
\label{table:effiency}
\end{table}

%% file: tabs/tradeoff.tex
\begin{table}[t]
\centering  
\tiny
\begin{tabularx}{\linewidth}{Xcccc}
\toprule
\textbf{Method} & \textbf{Avg. R@5}  & \textbf{Avg. R@10} & \textbf{Indexing Time } (h) $\downarrow$  \\
\midrule
Abstract & 66.9 & 72.1 & /   \\ 
PaperRegister (Ours) & \textbf{81.5} & \textbf{84.9} & 47.6   \\ 
~~w/o two-step registering& 76.1 & 78.9 & \textbf{14.5}   \\ 
\bottomrule
\end{tabularx}
\caption{PaperRegister can support sacrificing certain performance to alleviate offline time consumption.}
\label{table:tradeoff}
\end{table}

%% file: sections/app.tex
\newpage
\section{Detailed illustration for hierarchical register schema}
\label{app_a}

Since different types of papers have the different styles and formats, we design five kinds of hierarchical register schema, corresponding to five types of paper including algorithm innovation, benchmark construction, mechanism exploration, survey, and theory proof. We use a large language model to determine the type of input paper  and then assign the corresponding hierarchical register schema. The details of each kind of schema is in Figure~\ref{fig:app1}~\ref{fig:app2}~\ref{fig:app3}~\ref{fig:app4}~\ref{fig:app5}. And the instruction used to help LLM to do determine is 
\textit{
==Instruction==
Determine the category of the paper based on its abstract. The category options are as follows:  
Algorithm Innovation: Proposes new systems, new models, new training methods, new inference approaches, new data organization methods, etc.  
Benchmark Construction: Introduces a new benchmark.  
Mechanism Exploration: Investigates and analyzes the mechanisms of existing systems, algorithms, phenomena, or functionalities.  
Theory Proof: Proves a certain theory or formula.  
Survey and Review: Summarizes the research landscape of a particular field.  
Your final output should be only the best correct category of the paper, do not contain any other information or explanation.  
==Abstract==
abstract}

\section{Detailed process and instructions for fine-grained content extracting}
\label{app_b}

Learning from several widely recognized works, in which large language models are proven to possess reliable capabilities for content extraction~\cite{edge2024local,li2025deepsolution,li2024structrag,li2024meta}, PaperRegister uses a large language model as the extracting module. 
PaperRegister takes the node name and the text-formatted paper as input, uses instructions to guide in outputting corresponding content for the information node, and leaves it blank if the paper does not include corresponding content. In addition, PaperRegister also retrieve relevant paper text after the extracting as supplement to improve the register content.
The instruction used for LLM is 
\textit{==Instruction==  
You are an content extraction expert, particularly skilled at extracting structured records from academic papers. Below, I need you to perform extraction based on a given schema. Please note the following:
1. Your extraction does not need to preserve the original text verbatim. You may paraphrase or summarize the content to make the extracted content more comprehensive and fluent.  
2. Not all field names in the schema will have corresponding content in the paper. If you cannot find a precise match in the paper, leave that field empty.  
3. Your final output must strictly adhere to the original schema in JSON format, starting with '```json' and ending with '```'.  
==Schema==  
schema 
==Paper==  
paper}

\input{tabs/test_data}

\section{Detailed process and instructions for bottom-up content extracting}
\label{app_c}

PaperRegister takes contents of sub-nodes as input and uses instructions to guide the large language model in summarizing, condensing, and removing details to turn into summary text, thus obtaining the content for upper-layer information node. In addition, PaperRegister also retrieve relevant paper text after the extracting as supplement to improve the register content. The instruction used for LLM is 
\textit{==Instruction==  
You are an information integration expert, and now I need your help to complete an information integration task. 
I will provide you with a two-level tree structure, including a root node and two child nodes. Ideally, the content of the root node should be a summary and generalization of the two child nodes. Your task is to generate the content of the root node based on the content of the two child nodes. Note the following:  
1. The input I provide you is a dictionary in JSON format, including the keys `root\_name` and `children`, where the value of `children` is a list of child nodes.  
2. Each child node contains three fields: `node\_name`, `node\_desc`, and `node\_value`. You should primarily use the content of `node\_value` for summarization and generalization.  
3. The `root\_value` field should provide an abstraction and summary of the two child nodes’ contents. In other words, the root\_value must not repeat keywords from the child nodes; instead, it should abstract based on those keywords. More strictly, the root\_value length must not exceed that of either child node.
4. Your output should be a dictionary in JSON format, meaning the input dictionary will have content for the `root\_value` field. The final output should start with '```json' and end with '```'.  
==Input Tree==
tree
}

\section{Detailed statistical information of dataset for experiments}
\label{app_d}

We first use \textit{LitSearch},  mainly containing coarse-grained queries. And then we build a new dataset, \textit{Flexible-grained Search}, covering queries across various granularity and including data for test, development, and  training. 
Due to limited computational resources, we are unable to conduct experiments based on an ultra-large-scale paper corpus. For LitSearch, we extract 3,400 papers as the corpus, using the original 597 paper search queries. For Flexible-grained Search, we collect 4,200 papers as the corpus. The numbers of queries included in F.g.Search-1, F.g.Search-2, and F.g.Search-3 are 1,922, 1,922, and 1,800, respectively. The statics is in Table~\ref{tab:bench_statics}.

\section{Experimental results by more metrics}
\label{app_e}
To provide a more comprehensive illustration of the advancements of PaperRegister over baseline methods, we present experimental results in the Table~\ref{tab:precision_results},~\ref{tab:map_results},~\ref{tab:ndcg_results} across six metrics: Precision@5, Precision@10, MAP@5, MAP@10, NDCG@5, and NDCG@10. The data consistently support the conclusions drawn in our main experimental tables, further confirming that PaperRegister is an effective system for addressing flexible-grained paper search and the advantages of PaperRegister become more pronounced when handling finer-grained queries.

\input{figs/app1}

\input{figs/app2}

\input{figs/app3}

\input{figs/app4}

\input{figs/app5}

\input{tabs/main_pre}
\input{tabs/main_map}
\input{tabs/main_ndcg}

%% file: tabs/test_data.tex







\begin{table}[t]
\centering
\begin{tabular}{lcc}
\toprule
\textbf{Dataset Name} & \textbf{\# Corpus} & \textbf{\# Query} \\
\midrule
LitSearch & 3400 & 597 \\
F.g.Search-1 & 4200 & 1,922 \\
F.g.Search-2 & 4200 & 1,922 \\
F.g.Search-3 & 4200 & 1,800 \\
\bottomrule
\end{tabular}%
\caption{Statistics of test data in experiments.}
\label{tab:bench_statics}%
\end{table}%

%% file: figs/app1.tex
\begin{figure}[t!]
\centering
\includegraphics[width=\linewidth]{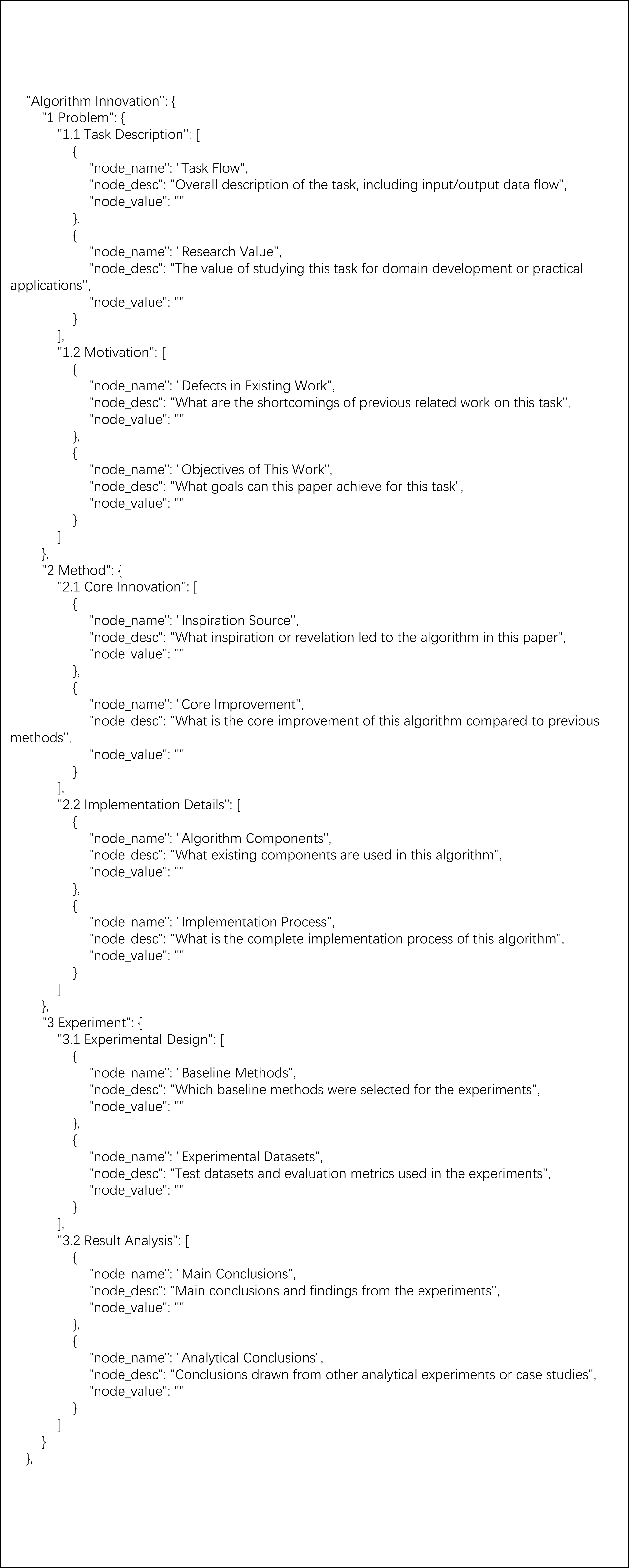} 
\caption{The first kind of hierarchical register schema (algorithm innovation).} 
\label{fig:app1}
\end{figure}

%% file: figs/app2.tex
\begin{figure}[t!]
\centering
\includegraphics[width=\linewidth]{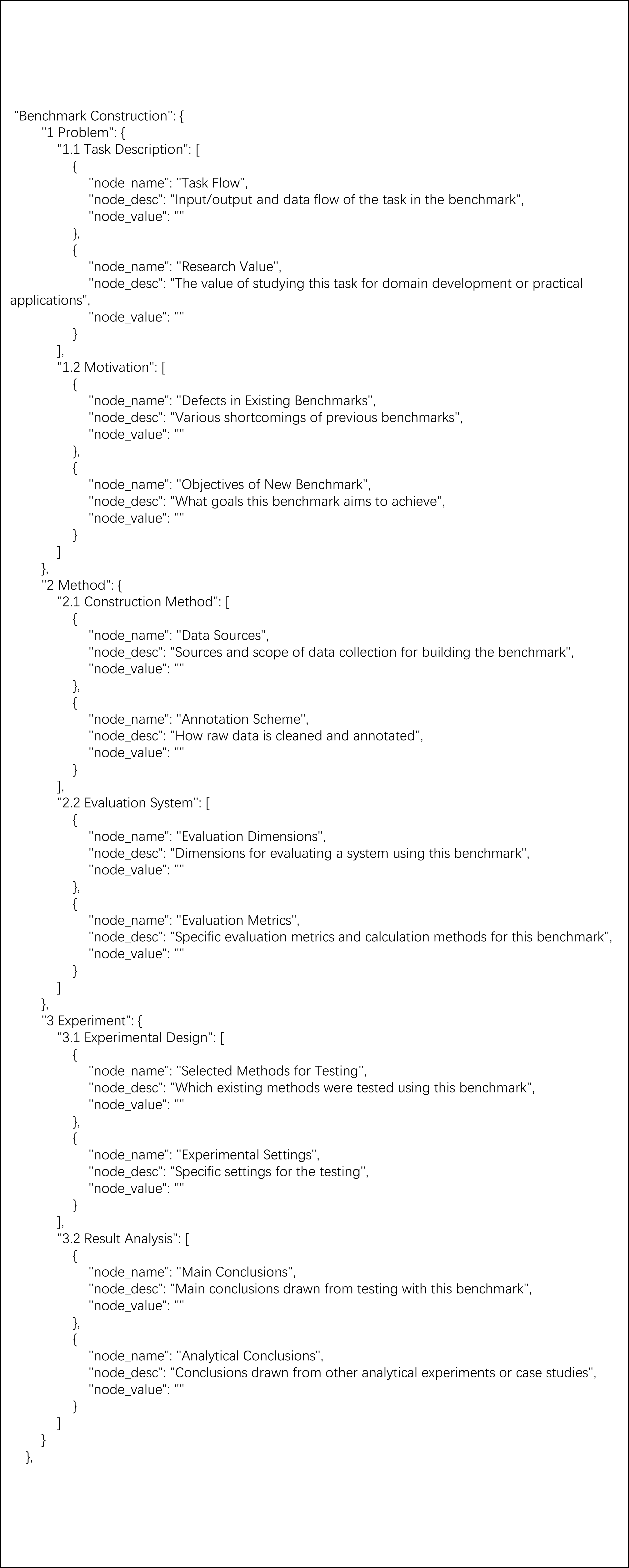} 
\caption{The second kind of hierarchical register schema (benchmark construction).} 
\label{fig:app2}
\end{figure}

%% file: figs/app3.tex
\begin{figure}[t!]
\centering
\includegraphics[width=\linewidth]{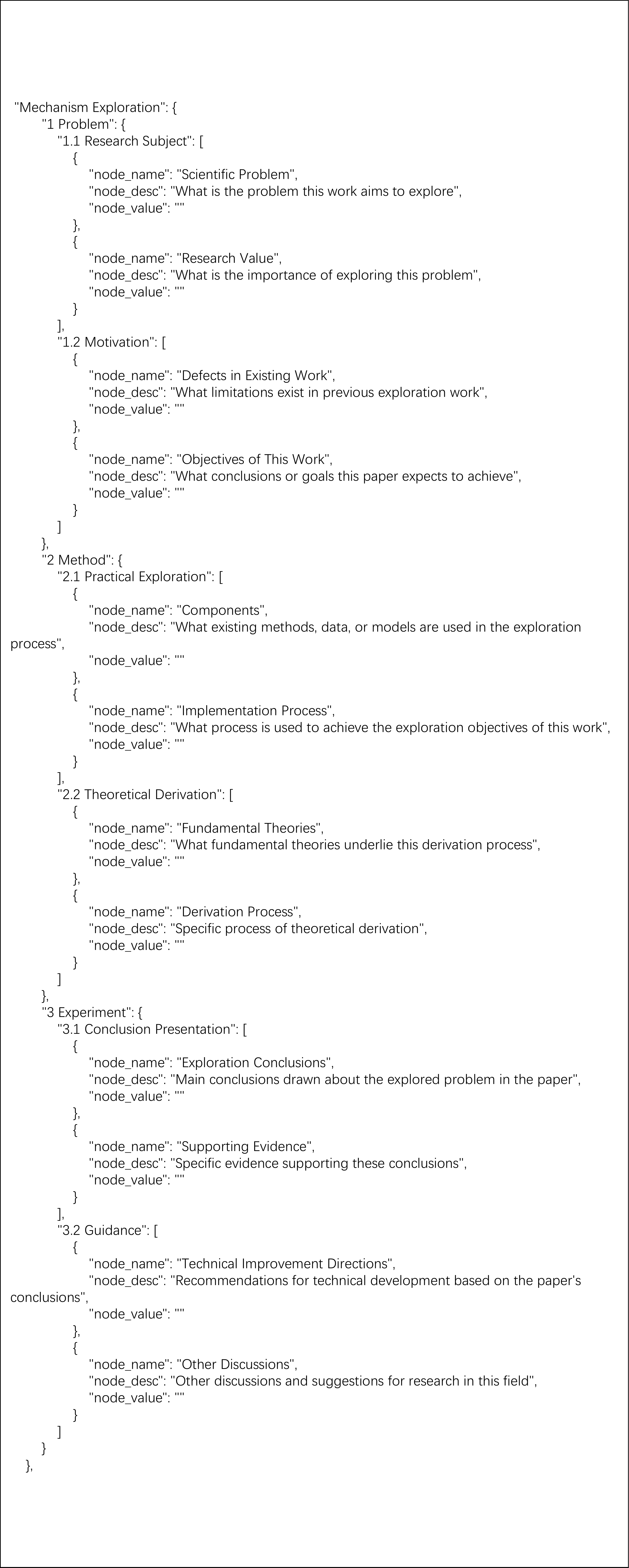} 
\caption{The third kind of hierarchical register schema (mechanism exploration).} 
\label{fig:app3}
\end{figure}

%% file: figs/app4.tex
\begin{figure}[t!]
\centering
\includegraphics[width=\linewidth]{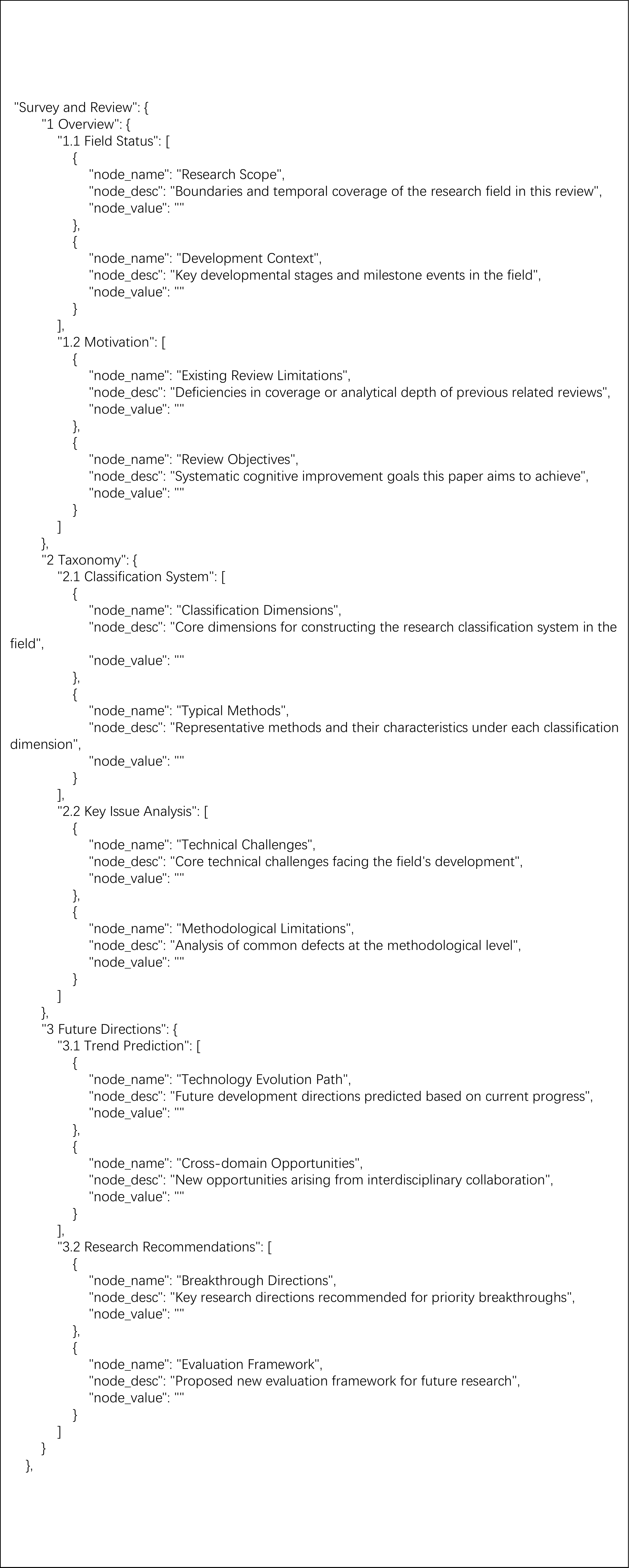} 
\caption{The fourth kind of hierarchical register schema (survey and review).} 
\label{fig:app4}
\end{figure}

%% file: figs/app5.tex
\begin{figure}[t!]
\centering
\includegraphics[width=\linewidth]{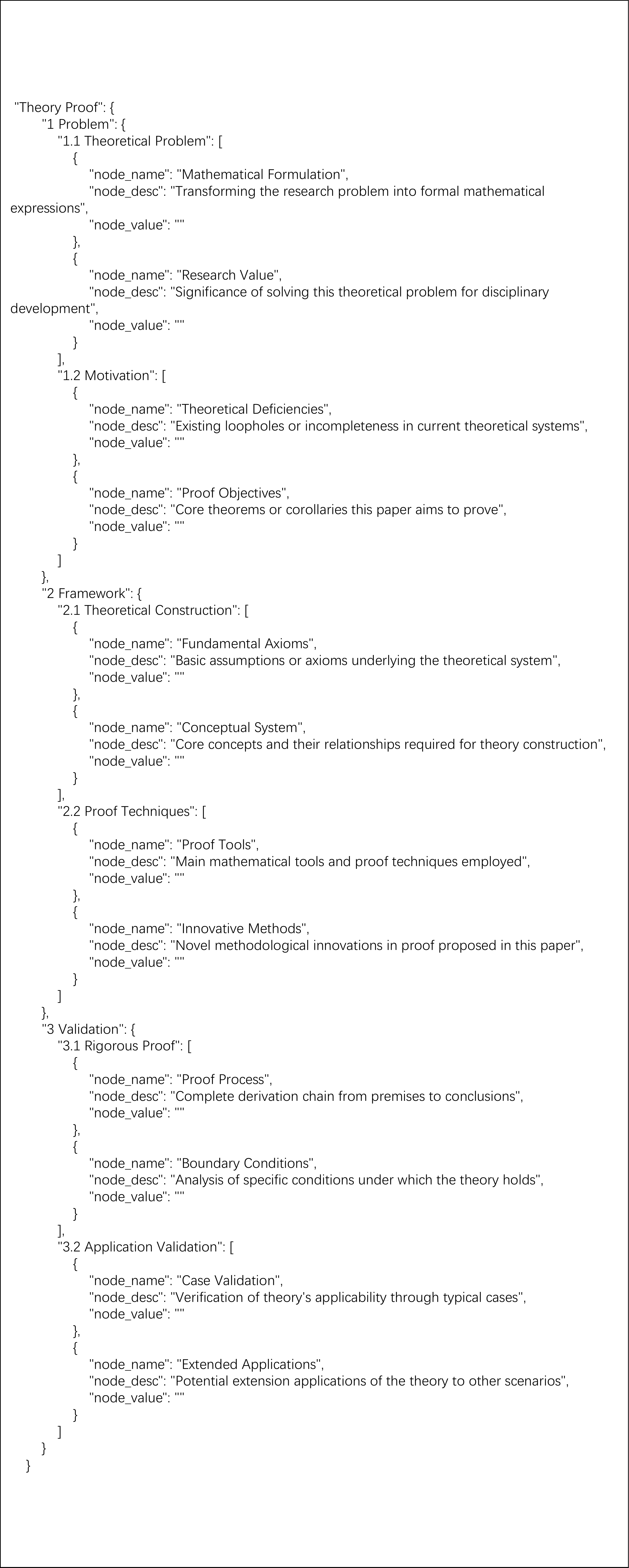} 
\caption{The fifth kind of hierarchical register schema (theory proof).} 
\label{fig:app5}
\end{figure}

%% file: tabs/main_pre.tex
\begin{table*}[t!]
\centering
\resizebox{\linewidth}{!}{
\begin{tabular}{lcccccccccccccccc}
\toprule

\multirow{4}[1]{*}{\textbf{Method}} & \multicolumn{8}{c}{\textbf{BM25-based Paper Search}} & \multicolumn{8}{c}{\textbf{DPR-based Paper Search}} \\

\cmidrule(lr){2-9} \cmidrule(lr){10-17}

& \multicolumn{2}{c}{\textbf{LitSearch}} & \multicolumn{2}{c}{\textbf{F.g.Search-1}} & \multicolumn{2}{c}{\textbf{F.g.Search-2}} & \multicolumn{2}{c}{\textbf{F.g.Search-3}} & \multicolumn{2}{c}{\textbf{LitSearch}} & \multicolumn{2}{c}{\textbf{F.g.Search-1}} & \multicolumn{2}{c}{\textbf{F.g.Search-2}} & \multicolumn{2}{c}{\textbf{F.g.Search-3}} \\

\cmidrule(lr){2-3} \cmidrule(lr){4-5}  \cmidrule(lr){6-7} \cmidrule(lr){8-9} \cmidrule(lr){10-11} \cmidrule(lr){12-13}  \cmidrule(lr){14-15} \cmidrule(lr){16-17}
& P@5 & P@10 & P@5 & P@10 & P@5 & P@10 & P@5 & P@10 & P@5 & P@10 & P@5 & P@10 & P@5 & P@10 & P@5 & P@10 \\

\midrule
\rowcolor[rgb]{ .906,  .902,  .902} \multicolumn{17}{c}{Direct Matching} \\
\midrule
Title                                              & 11.4          & 6.3          & 8.4          & 4.7          & 7.2          & 4.2          & 6.1          & 3.5          & 14.0          & 7.8          & 10.4          & 5.9          & 9.2          & 5.2          & 8.1          & 4.7          \\
Abstract                                           & 14.1          & 7.6          & 12.7          & 6.9          & 11.5          & 6.3          & 10.8          & 6.0          & 16.4          & 8.8          & 13.9          & 7.4          & 12.4          & 6.8          & 11.6          & 6.3          \\
Total Paper                                        & 13.5          & 7.3          & 15.9          & 8.2          & 16.3          & 8.4          & 16.8          & 8.6          & 16.0          & 8.8          & 14.6          & 7.9          & 13.8          & 7.4          & 13.2          & 7.1          \\
\midrule
\rowcolor[rgb]{ .906,  .902,  .902} \multicolumn{17}{c}{Query Enhancing} \\
\midrule
Rewriting~\cite{ma2023query}
& 13.0          & 7.4          & 11.7          & 6.4          & 10.8          & 6.0          & 10.2          & 5.6          & 16.0          & 8.8          & 13.4          & 7.3          & 12.1          & 6.5          & 11.0          & 6.1          \\
HyDE~\cite{gao2023precise}
& \textbf{14.6}          & \textbf{8.0}          & 12.1          & 6.6          & 10.9          & 6.0          & 10.3          & 5.8          & 16.2          & 8.9          & 13.4          & 7.2          & 11.8          & 6.6          & 11.3          & 6.2          \\
CSQE~\cite{lei2024corpus}
& 14.4          & 7.8          & 11.8          & 6.4          & 10.8          & 5.9          & 10.3          & 5.6          & 16.5          & 8.7          & 13.2          & 7.1          & 12.0          & 6.5          & 11.0          & 6.1          \\

\midrule
\rowcolor[rgb]{ .906,  .902,  .902} \multicolumn{17}{c}{Multi-field Indexing} \\
\midrule

$\text{Chunk}_{avg}$~\cite{li2025multifield}                               & 12.2          & 7.2          & 13.6          & 7.4          & 13.3          & 7.2          & 13.2          & 7.2          & 10.5          & 6.2          & 10.2          & 5.8          & 9.3          & 5.3          & 9.3          & 5.2          \\
$\text{Chunk}_{max}$~\cite{li2025multifield}                               & 14.3          & 7.9          & 16.0          & 8.3          & 16.3          & 8.5          & 17.1          & 8.8          & 15.2          & 8.5          & 16.0          & 8.3          & 15.3          & 8.1          & 15.8          & 8.2          \\
$\text{Paragraph}_{avg}$~\cite{li2025multifield}                           & 6.2          & 4.1          & 11.8          & 6.7          & 12.0          & 6.7          & 11.6          & 6.6          & 4.3          & 2.4          & 4.7          & 3.1          & 4.6          & 3.0          & 4.5          & 2.9          \\
$\text{Paragraph}_{max}$~\cite{li2025multifield}                           & 13.5          & 7.5          & 14.8          & 7.9          & 15.1          & 8.0          & 16.2          & 8.4          & 16.7          & 9.0          & 15.8          & 8.3          & 15.3          & 8.1          & 15.8          & 8.2          \\

\midrule
\rowcolor[rgb]{ .906,  .902,  .902} \multicolumn{17}{c}{Hierarchical Register Indexing} \\
\midrule

PaperRegister (Ours)                               & \textbf{14.6} & \textbf{8.0} & \textbf{17.9} & \textbf{9.1} & \textbf{17.6} & \textbf{8.9} & \textbf{17.5} & \textbf{8.9} & \textbf{17.1} & \textbf{9.3} & \textbf{16.8} & \textbf{8.7} & \textbf{16.0} & \textbf{8.3} & \textbf{16.2} & \textbf{8.3} \\

\bottomrule
\end{tabular}%
}
\caption{Precision@5 and Precision@10 results on paper search across various granularity.}
\label{tab:precision_results}%
\end{table*}%

%% file: tabs/main_map.tex
\begin{table*}[t!]
\centering
\resizebox{\linewidth}{!}{
\begin{tabular}{lcccccccccccccccc}
\toprule

\multirow{4}[1]{*}{\textbf{Method}} & \multicolumn{8}{c}{\textbf{BM25-based Paper Search}} & \multicolumn{8}{c}{\textbf{DPR-based Paper Search}} \\

\cmidrule(lr){2-9} \cmidrule(lr){10-17}

& \multicolumn{2}{c}{\textbf{LitSearch}} & \multicolumn{2}{c}{\textbf{F.g.Search-1}} & \multicolumn{2}{c}{\textbf{F.g.Search-2}} & \multicolumn{2}{c}{\textbf{F.g.Search-3}} & \multicolumn{2}{c}{\textbf{LitSearch}} & \multicolumn{2}{c}{\textbf{F.g.Search-1}} & \multicolumn{2}{c}{\textbf{F.g.Search-2}} & \multicolumn{2}{c}{\textbf{F.g.Search-3}} \\

\cmidrule(lr){2-3} \cmidrule(lr){4-5}  \cmidrule(lr){6-7} \cmidrule(lr){8-9} \cmidrule(lr){10-11} \cmidrule(lr){12-13}  \cmidrule(lr){14-15} \cmidrule(lr){16-17}
& M@5 & M@10 & M@5 & M@10 & M@5 & M@10 & M@5 & M@10 & M@5 & M@10 & M@5 & M@10 & M@5 & M@10 & M@5 & M@10 \\

\midrule
\rowcolor[rgb]{ .906,  .902,  .902} \multicolumn{17}{c}{Direct Matching} \\
\midrule
Title                                              & 42.2          & 42.9          & 33.4          & 34.1          & 28.8          & 29.6          & 23.5          & 24.2          & 54.9          & 55.9          & 41.3          & 42.2          & 36.3          & 37.2          & 32.2          & 33.0          \\
Abstract                                           & 52.4          & 53.0          & 54.9          & 55.6          & 49.4          & 50.1          & 45.8          & 46.5          & 65.1          & 65.9          & 59.8          & 60.5          & 51.9          & 52.7          & 47.2          & 47.8          \\
Total Paper                                        & 51.7          & 52.4          & 74.3          & 74.6          & 76.1          & 76.4          & 79.1          & 79.3          & 60.5          & 61.6          & 63.2          & 64.0          & 59.2          & 59.9          & 55.6          & 56.3          \\
\midrule
\rowcolor[rgb]{ .906,  .902,  .902} \multicolumn{17}{c}{Query Enhancing} \\
\midrule
Rewriting~\cite{ma2023query}
& 47.2          & 48.3          & 48.7          & 49.5          & 44.8          & 45.5          & 41.2          & 41.9          & 62.3          & 63.4          & 56.7          & 57.4          & 50.1          & 50.8          & 44.8          & 45.6          \\
HyDE~\cite{gao2023precise}
& 54.1          & 55.1          & 50.1          & 50.9          & 44.6          & 45.3          & 41.7          & 42.6          & 64.3          & 65.3          & 56.4          & 57.1          & 48.7          & 49.6          & 45.1          & 45.9          \\
CSQE~\cite{lei2024corpus}
& 54.1          & 54.8          & 41.2          & 42.0          & 37.8          & 38.4          & 36.7          & 37.4          & 65.2          & 65.8          & 54.1          & 54.9          & 48.4          & 49.1          & 44.4          & 45.1          \\

\midrule
\rowcolor[rgb]{ .906,  .902,  .902} \multicolumn{17}{c}{Multi-field Indexing} \\
\midrule

$\text{Chunk}_{avg}$~\cite{li2025multifield}                               & 40.6          & 42.0          & 54.4          & 55.2          & 54.9          & 55.6          & 54.4          & 55.2          & 34.9          & 36.1          & 39.4          & 40.4          & 36.6          & 37.5          & 35.1          & 35.9          \\
$\text{Chunk}_{max}$~\cite{li2025multifield}                               & 56.1          & 57.1          & 73.6          & 74.0          & 75.3          & 75.8          & 79.5          & 79.8          & 62.0          & 63.0          & 71.7          & 72.2          & 68.4          & 68.9          & 69.4          & 69.9          \\
$\text{Paragraph}_{avg}$~\cite{li2025multifield}                           & 19.0          & 20.2          & 44.9          & 46.0          & 46.2          & 47.1          & 44.8          & 45.9          & 15.0          & 15.3          & 15.9          & 16.8          & 15.7          & 16.6          & 15.1          & 15.9          \\
$\text{Paragraph}_{max}$~\cite{li2025multifield}                           & 50.7          & 51.7          & 66.4          & 67.0          & 67.2          & 67.8          & 73.2          & 73.6          & 66.1          & 67.0          & 70.8          & 71.3          & 68.5          & 69.1          & 68.7          & 69.1          \\

\midrule
\rowcolor[rgb]{ .906,  .902,  .902} \multicolumn{17}{c}{Hierarchical Register Indexing} \\
\midrule

PaperRegister (Ours)                               & \textbf{57.4} & \textbf{58.3} & \textbf{86.2} & \textbf{86.4} & \textbf{84.9} & \textbf{85.0} & \textbf{85.3} & \textbf{85.5} & \textbf{70.0} & \textbf{70.9} & \textbf{77.4} & \textbf{77.8} & \textbf{73.7} & \textbf{74.0} & \textbf{72.9} & \textbf{73.1} \\

\bottomrule
\end{tabular}%
}
\caption{MAP@5 and MAP@10 results on paper search across various granularity.}
\label{tab:map_results}%
\end{table*}%

%% file: tabs/main_ndcg.tex
\begin{table*}[t!]
\centering
\resizebox{\linewidth}{!}{
\begin{tabular}{lcccccccccccccccc}
\toprule

\multirow{4}[1]{*}{\textbf{Method}} & \multicolumn{8}{c}{\textbf{BM25-based Paper Search}} & \multicolumn{8}{c}{\textbf{DPR-based Paper Search}} \\

\cmidrule(lr){2-9} \cmidrule(lr){10-17}

& \multicolumn{2}{c}{\textbf{LitSearch}} & \multicolumn{2}{c}{\textbf{F.g.Search-1}} & \multicolumn{2}{c}{\textbf{F.g.Search-2}} & \multicolumn{2}{c}{\textbf{F.g.Search-3}} & \multicolumn{2}{c}{\textbf{LitSearch}} & \multicolumn{2}{c}{\textbf{F.g.Search-1}} & \multicolumn{2}{c}{\textbf{F.g.Search-2}} & \multicolumn{2}{c}{\textbf{F.g.Search-3}} \\

\cmidrule(lr){2-3} \cmidrule(lr){4-5}  \cmidrule(lr){6-7} \cmidrule(lr){8-9} \cmidrule(lr){10-11} \cmidrule(lr){12-13}  \cmidrule(lr){14-15} \cmidrule(lr){16-17}
& N@5 & N@10 & N@5 & N@10 & N@5 & N@10 & N@5 & N@10 & N@5 & N@10 & N@5 & N@10 & N@5 & N@10 & N@5 & N@10 \\

\midrule
\rowcolor[rgb]{ .906,  .902,  .902} \multicolumn{17}{c}{Direct Matching} \\
\midrule
Title                                              & 45.4          & 47.1          & 35.6          & 37.3          & 30.6          & 32.5          & 25.2          & 26.8          & 58.2          & 60.3          & 44.0          & 46.2          & 38.7          & 40.7          & 34.3          & 36.4          \\
Abstract                                           & 56.3          & 57.7          & 57.1          & 58.8          & 51.4          & 53.1          & 47.9          & 49.6          & 68.5          & 70.3          & 62.2          & 63.8          & 54.5          & 56.4          & 50.0          & 51.5          \\
Total Paper                                        & 55.0          & 56.6          & 75.6          & 76.5          & 77.5          & 78.1          & 80.3          & 81.0          & 64.4          & 66.8          & 65.7          & 67.5          & 61.6          & 63.3          & 58.2          & 59.8          \\
\midrule
\rowcolor[rgb]{ .906,  .902,  .902} \multicolumn{17}{c}{Query Enhancing} \\
\midrule
Rewriting~\cite{ma2023query}
& 51.2          & 53.7          & 51.1          & 53.0          & 47.1          & 48.9          & 43.6          & 45.4          & 66.0          & 68.4          & 59.3          & 61.1          & 52.7          & 54.3          & 47.4          & 49.3          \\
HyDE~\cite{gao2023precise}
& 58.1          & 60.3          & 52.7          & 54.6          & 47.0          & 48.9          & 44.2          & 46.1          & 67.6          & 69.9          & 59.0          & 60.8          & 51.3          & 53.5          & 47.9          & 49.8          \\
CSQE~\cite{lei2024corpus}
& 58.0          & 59.7          & 45.7          & 47.4          & 41.9          & 43.4          & 40.4          & 41.9          & 68.6          & 69.8          & 57.1          & 58.8          & 51.4          & 53.0          & 47.1          & 48.9          \\

\midrule
\rowcolor[rgb]{ .906,  .902,  .902} \multicolumn{17}{c}{Multi-field Indexing} \\
\midrule

$\text{Chunk}_{avg}$~\cite{li2025multifield}                               & 45.2          & 48.4          & 57.8          & 59.8          & 57.8          & 59.5          & 57.3          & 59.1          & 38.8          & 41.8          & 42.3          & 44.7          & 39.1          & 41.3          & 37.9          & 39.8          \\
$\text{Chunk}_{max}$~\cite{li2025multifield}                               & 59.2          & 61.7          & 75.2          & 76.2          & 76.8          & 77.9          & 81.0          & 81.7          & 64.9          & 67.3          & 73.8          & 74.8          & 70.5          & 71.7          & 71.8          & 72.8          \\
$\text{Paragraph}_{avg}$~\cite{li2025multifield}                           & 21.8          & 24.7          & 48.4          & 50.9          & 49.6          & 51.7          & 48.2          & 50.8          & 16.4          & 17.2          & 17.7          & 20.0          & 17.5          & 19.7          & 16.9          & 19.0          \\
$\text{Paragraph}_{max}$~\cite{li2025multifield}                           & 54.4          & 56.6          & 68.2          & 69.9          & 69.3          & 70.7          & 75.1          & 76.1          & 69.7          & 71.5          & 72.9          & 74.1          & 70.5          & 72.1          & 71.2          & 72.2          \\

\midrule
\rowcolor[rgb]{ .906,  .902,  .902} \multicolumn{17}{c}{Hierarchical Register Indexing} \\
\midrule

PaperRegister (Ours)                               & \textbf{60.7} & \textbf{62.9} & \textbf{87.1} & \textbf{87.5} & \textbf{85.7} & \textbf{86.0} & \textbf{85.9} & \textbf{86.3} & \textbf{73.0} & \textbf{75.0} & \textbf{79.1} & \textbf{80.1} & \textbf{75.2} & \textbf{76.1} & \textbf{74.9} & \textbf{75.5} \\

\bottomrule
\end{tabular}%
}
\caption{NDCG@5 and NDCG@10 results on paper search across various granularity.}
\label{tab:ndcg_results}%
\end{table*}%

%% file: custom.bib
@article{Shao2024DeepSeekMathPT,
 author = {Zhihong Shao and Peiyi Wang and Qihao Zhu and Runxin Xu and Jun-Mei Song and Mingchuan Zhang and Y. K. Li and Yu Wu and Daya Guo},
 journal = {ArXiv preprint},
 title = {DeepSeekMath: Pushing the Limits of Mathematical Reasoning in Open Language Models},
 url = {https://arxiv.org/abs/2402.03300},
 volume = {abs/2402.03300},
 year = {2024}
}

@article{kuhlthau1991inside,
 author = {Kuhlthau, Carol C},
 journal = {Journal of the American society for information science},
 number = {5},
 pages = {361--371},
 publisher = {Wiley Online Library},
 title = {Inside the search process: Information seeking from the user's perspective},
 volume = {42},
 year = {1991}
}

@article{ellis1993comparison,
 author = {Ellis, David and Cox, Deborah and Hall, Katherine},
 journal = {Journal of documentation},
 number = {4},
 pages = {356--369},
 publisher = {MCB UP Ltd},
 title = {A comparison of the information seeking patterns of researchers in the physical and social sciences},
 volume = {49},
 year = {1993}
}

@article{hemminger2007information,
 author = {Hemminger, Bradley M and Lu, Dihui and Vaughan, KTL and Adams, Stephanie J},
 journal = {Journal of the American society for information science and technology},
 number = {14},
 pages = {2205--2225},
 publisher = {Wiley Online Library},
 title = {Information seeking behavior of academic scientists},
 volume = {58},
 year = {2007}
}

@book{case2016looking,
 author = {Case, Donald O and Given, Lisa M},
 publisher = {Emerald Group Publishing},
 title = {Looking for information: A survey of research on information seeking, needs, and behavior},
 year = {2016}
}

@inproceedings{cohan2020specter,
 address = {Online},
 author = {Cohan, Arman  and
Feldman, Sergey  and
Beltagy, Iz  and
Downey, Doug  and
Weld, Daniel},
 booktitle = {Proceedings of the 58th Annual Meeting of the Association for Computational Linguistics},
 doi = {10.18653/v1/2020.acl-main.207},
 editor = {Jurafsky, Dan  and
Chai, Joyce  and
Schluter, Natalie  and
Tetreault, Joel},
 pages = {2270--2282},
 publisher = {Association for Computational Linguistics},
 title = {{SPECTER}: Document-level Representation Learning using Citation-informed Transformers},
 url = {https://aclanthology.org/2020.acl-main.207},
 year = {2020}
}

@inproceedings{wadden2020fact,
 address = {Online},
 author = {Wadden, David  and
Lin, Shanchuan  and
Lo, Kyle  and
Wang, Lucy Lu  and
van Zuylen, Madeleine  and
Cohan, Arman  and
Hajishirzi, Hannaneh},
 booktitle = {Proceedings of the 2020 Conference on Empirical Methods in Natural Language Processing (EMNLP)},
 doi = {10.18653/v1/2020.emnlp-main.609},
 editor = {Webber, Bonnie  and
Cohn, Trevor  and
He, Yulan  and
Liu, Yang},
 pages = {7534--7550},
 publisher = {Association for Computational Linguistics},
 title = {Fact or Fiction: Verifying Scientific Claims},
 url = {https://aclanthology.org/2020.emnlp-main.609},
 year = {2020}
}

@inproceedings{wang2023scientific,
 author = {Jianyou Wang and
Kaicheng Wang and
Xiaoyue Wang and
Prudhviraj Naidu and
Leon Bergen and
Ramamohan Paturi},
 bibsource = {dblp computer science bibliography, https://dblp.org},
 booktitle = {Advances in Neural Information Processing Systems 36: Annual Conference
on Neural Information Processing Systems 2023, NeurIPS 2023, New Orleans,
LA, USA, December 10 - 16, 2023},
 editor = {Alice Oh and
Tristan Naumann and
Amir Globerson and
Kate Saenko and
Moritz Hardt and
Sergey Levine},
 title = {Scientific Document Retrieval using Multi-level Aspect-based Queries},
 url = {http://papers.nips.cc/paper\_files/paper/2023/hash/78f9c04bdcb06f1ada3902912d8b64ba-Abstract-Datasets\_and\_Benchmarks.html},
 year = {2023}
}

@article{zhang2025scientific,
 author = {Zhang, Yunyi and Yang, Ruozhen and Jiao, Siqi and Kang, SeongKu and Han, Jiawei},
 journal = {ArXiv preprint},
 title = {Scientific Paper Retrieval with LLM-Guided Semantic-Based Ranking},
 url = {https://arxiv.org/abs/2505.21815},
 volume = {abs/2505.21815},
 year = {2025}
}

@inproceedings{zheng2020bert,
 address = {Online},
 author = {Zheng, Zhi  and
Hui, Kai  and
He, Ben  and
Han, Xianpei  and
Sun, Le  and
Yates, Andrew},
 booktitle = {Findings of the Association for Computational Linguistics: EMNLP 2020},
 doi = {10.18653/v1/2020.findings-emnlp.424},
 editor = {Cohn, Trevor  and
He, Yulan  and
Liu, Yang},
 pages = {4718--4728},
 publisher = {Association for Computational Linguistics},
 title = {{BERT-QE}: {C}ontextualized {Q}uery {E}xpansion for {D}ocument {R}e-ranking},
 url = {https://aclanthology.org/2020.findings-emnlp.424},
 year = {2020}
}

@inproceedings{mackie2023generative,
 author = {Iain Mackie and
Shubham Chatterjee and
Jeffrey Dalton},
 bibsource = {dblp computer science bibliography, https://dblp.org},
 booktitle = {Proceedings of the 46th International {ACM} {SIGIR} Conference on
Research and Development in Information Retrieval, {SIGIR} 2023, Taipei,
Taiwan, July 23-27, 2023},
 doi = {10.1145/3539618.3591992},
 editor = {Hsin{-}Hsi Chen and
Wei{-}Jou (Edward) Duh and
Hen{-}Hsen Huang and
Makoto P. Kato and
Josiane Mothe and
Barbara Poblete},
 pages = {2026--2031},
 publisher = {{ACM}},
 title = {Generative Relevance Feedback with Large Language Models},
 url = {https://doi.org/10.1145/3539618.3591992},
 year = {2023}
}

@inproceedings{lei2024corpus,
 address = {St. Julian{'}s, Malta},
 author = {Lei, Yibin  and
Cao, Yu  and
Zhou, Tianyi  and
Shen, Tao  and
Yates, Andrew},
 booktitle = {Proceedings of the 18th Conference of the European Chapter of the Association for Computational Linguistics (Volume 2: Short Papers)},
 editor = {Graham, Yvette  and
Purver, Matthew},
 pages = {393--401},
 publisher = {Association for Computational Linguistics},
 title = {Corpus-Steered Query Expansion with Large Language Models},
 url = {https://aclanthology.org/2024.eacl-short.34},
 year = {2024}
}

@article{li2024meta,
  title={Meta-Cognitive Analysis: Evaluating Declarative and Procedural Knowledge in Datasets and Large Language Models},
  author={Li, Zhuoqun and Lin, Hongyu and Lu, Yaojie and Xiang, Hao and Han, Xianpei and Sun, Le},
  journal={arXiv preprint arXiv:2403.09750},
  year={2024}
}

@article{li2024structrag,
 author = {Li, Zhuoqun and Chen, Xuanang and Yu, Haiyang and Lin, Hongyu and Lu, Yaojie and Tang, Qiaoyu and Huang, Fei and Han, Xianpei and Sun, Le and Li, Yongbin},
 journal = {ArXiv preprint},
 title = {Structrag: Boosting knowledge intensive reasoning of llms via inference-time hybrid information structurization},
 url = {https://arxiv.org/abs/2410.08815},
 volume = {abs/2410.08815},
 year = {2024}
}

@article{edge2024local,
 author = {Edge, Darren and Trinh, Ha and Cheng, Newman and Bradley, Joshua and Chao, Alex and Mody, Apurva and Truitt, Steven and Metropolitansky, Dasha and Ness, Robert Osazuwa and Larson, Jonathan},
 journal = {ArXiv preprint},
 title = {From local to global: A graph rag approach to query-focused summarization},
 url = {https://arxiv.org/abs/2404.16130},
 volume = {abs/2404.16130},
 year = {2024}
}

@inproceedings{Holtzman2020The,
 author = {Ari Holtzman and
Jan Buys and
Li Du and
Maxwell Forbes and
Yejin Choi},
 bibsource = {dblp computer science bibliography, https://dblp.org},
 booktitle = {8th International Conference on Learning Representations, {ICLR} 2020,
Addis Ababa, Ethiopia, April 26-30, 2020},
 publisher = {OpenReview.net},
 title = {The Curious Case of Neural Text Degeneration},
 url = {https://openreview.net/forum?id=rygGQyrFvH},
 year = {2020}
}

@article{qwen3,
 author = {Team Qwen},
 journal = {ArXiv preprint},
 title = {Qwen3 Technical Report},
 url = {https://arxiv.org/abs/2505.09388},
 volume = {abs/2505.09388},
 year = {2025}
}

@article{li2023towards,
 author = {Li, Zehan and Zhang, Xin and Zhang, Yanzhao and Long, Dingkun and Xie, Pengjun and Zhang, Meishan},
 journal = {ArXiv preprint},
 title = {Towards general text embeddings with multi-stage contrastive learning},
 url = {https://arxiv.org/abs/2308.03281},
 volume = {abs/2308.03281},
 year = {2023}
}

@inproceedings{tang2024selfretrieval,
 author = {Qiaoyu Tang and
Jiawei Chen and
Zhuoqun Li and
Bowen Yu and
Yaojie Lu and
Cheng Fu and
Haiyang Yu and
Hongyu Lin and
Fei Huang and
Ben He and
Xianpei Han and
Le Sun and
Yongbin Li},
 bibsource = {dblp computer science bibliography, https://dblp.org},
 booktitle = {Advances in Neural Information Processing Systems 38: Annual Conference
on Neural Information Processing Systems 2024, NeurIPS 2024, Vancouver,
BC, Canada, December 10 - 15, 2024},
 editor = {Amir Globersons and
Lester Mackey and
Danielle Belgrave and
Angela Fan and
Ulrich Paquet and
Jakub M. Tomczak and
Cheng Zhang},
 title = {Self-Retrieval: End-to-End Information Retrieval with One Large Language
Model},
 url = {http://papers.nips.cc/paper\_files/paper/2024/hash/741ad162ab0f3da6f9aad60e9e34f5f1-Abstract-Conference.html},
 year = {2024}
}

@inproceedings{ma2023query,
 address = {Singapore},
 author = {Ma, Xinbei  and
Gong, Yeyun  and
He, Pengcheng  and
Zhao, Hai  and
Duan, Nan},
 booktitle = {Proceedings of the 2023 Conference on Empirical Methods in Natural Language Processing},
 doi = {10.18653/v1/2023.emnlp-main.322},
 editor = {Bouamor, Houda  and
Pino, Juan  and
Bali, Kalika},
 pages = {5303--5315},
 publisher = {Association for Computational Linguistics},
 title = {Query Rewriting in Retrieval-Augmented Large Language Models},
 url = {https://aclanthology.org/2023.emnlp-main.322},
 year = {2023}
}

@inproceedings{gao2023precise,
 address = {Toronto, Canada},
 author = {Gao, Luyu  and
Ma, Xueguang  and
Lin, Jimmy  and
Callan, Jamie},
 booktitle = {Proceedings of the 61st Annual Meeting of the Association for Computational Linguistics (Volume 1: Long Papers)},
 doi = {10.18653/v1/2023.acl-long.99},
 editor = {Rogers, Anna  and
Boyd-Graber, Jordan  and
Okazaki, Naoaki},
 pages = {1762--1777},
 publisher = {Association for Computational Linguistics},
 title = {Precise Zero-Shot Dense Retrieval without Relevance Labels},
 url = {https://aclanthology.org/2023.acl-long.99},
 year = {2023}
}

@inproceedings{takeshita-etal-2024-gengo,
    title = "{G}en{GO}: {ACL} Paper Explorer with Semantic Features",
    author = {Takeshita Sotaro  and
      Ponzetto Simone  and
      Eckert Kai and Takeshita Sotaro},
    editor = "Cao, Yixin  and
      Feng, Yang  and
      Xiong, Deyi",
    booktitle = "Proceedings of the 62nd Annual Meeting of the Association for Computational Linguistics (Volume 3: System Demonstrations)",
    month = aug,
    year = "2024",
    address = "Bangkok, Thailand",
    publisher = "Association for Computational Linguistics",
    url = "https://aclanthology.org/2024.acl-demos.12/",
    doi = "10.18653/v1/2024.acl-demos.12",
    pages = "117--126"
}

@inproceedings{
li2025multifield,
title={Multi-Field Adaptive Retrieval},
author={Millicent Li and Tongfei Chen and Benjamin Van Durme and Patrick Xia},
booktitle={The Thirteenth International Conference on Learning Representations},
year={2025},
url={https://openreview.net/forum?id=3PDklqqqfN}
}

@article{shi2025hypercube,
  title={Hypercube-RAG: Hypercube-Based Retrieval-Augmented Generation for In-domain Scientific Question-Answering},
  author={Shi, Jimeng and Zhou, Sizhe and Jin, Bowen and Hu, Wei and Wang, Shaowen and Narasimhan, Giri and Han, Jiawei},
  journal={arXiv preprint arXiv:2505.19288},
  year={2025}
}

@article{chen2025enrichindex,
  title={EnrichIndex: Using LLMs to Enrich Retrieval Indices Offline},
  author={Chen, Peter Baile and Wolfson, Tomer and Cafarella, Michael and Roth, Dan},
  journal={arXiv preprint arXiv:2504.03598},
  year={2025}
}

@article{anand2023context,
  title={Context aware query rewriting for text rankers using llm},
  author={Anand, Abhijit and Setty, Vinay and Anand, Avishek and others},
  journal={arXiv preprint arXiv:2308.16753},
  year={2023}
}

@article{li2024automir,
  title={AutoMIR: Effective Zero-Shot Medical Information Retrieval without Relevance Labels},
  author={Li, Lei and Zhang, Xiangxu and Zhou, Xiao and Liu, Zheng},
  journal={arXiv preprint arXiv:2410.20050},
  year={2024}
}

@article{ren2025towards,
  title={Towards scientific intelligence: A survey of llm-based scientific agents},
  author={Ren, Shuo and Jian, Pu and Ren, Zhenjiang and Leng, Chunlin and Xie, Can and Zhang, Jiajun},
  journal={arXiv preprint arXiv:2503.24047},
  year={2025}
}

@article{li2025pseudo,
  title={Pseudo Relevance Feedback is Enough to Close the Gap Between Small and Large Dense Retrieval Models},
  author={Li, Hang and Wang, Xiao and Koopman, Bevan and Zuccon, Guido},
  journal={arXiv preprint arXiv:2503.14887},
  year={2025}
}

@inproceedings{li2025deepsolution,
    title = "{D}eep{S}olution: Boosting Complex Engineering Solution Design via Tree-based Exploration and Bi-point Thinking",
    author = "Li, Zhuoqun  and
      Yu, Haiyang  and
      Chen, Xuanang  and
      Lin, Hongyu  and
      Lu, Yaojie  and
      Huang, Fei  and
      Han, Xianpei  and
      Li, Yongbin  and
      Sun, Le",
    editor = "Che, Wanxiang  and
      Nabende, Joyce  and
      Shutova, Ekaterina  and
      Pilehvar, Mohammad Taher",
    booktitle = "Proceedings of the 63rd Annual Meeting of the Association for Computational Linguistics (Volume 1: Long Papers)",
    month = jul,
    year = "2025",
    address = "Vienna, Austria",
    publisher = "Association for Computational Linguistics",
    url = "https://aclanthology.org/2025.acl-long.220/",
    pages = "4380--4396",
    ISBN = "979-8-89176-251-0"
}

@inproceedings{he2025pasa,
    title = "{P}a{S}a: An {LLM} Agent for Comprehensive Academic Paper Search",
    author = "He, Yichen  and
      Huang, Guanhua  and
      Feng, Peiyuan  and
      Lin, Yuan  and
      Zhang, Yuchen  and
      Li, Hang  and
      E, Weinan",
    editor = "Che, Wanxiang  and
      Nabende, Joyce  and
      Shutova, Ekaterina  and
      Pilehvar, Mohammad Taher",
    booktitle = "Proceedings of the 63rd Annual Meeting of the Association for Computational Linguistics (Volume 1: Long Papers)",
    month = jul,
    year = "2025",
    address = "Vienna, Austria",
    publisher = "Association for Computational Linguistics",
    url = "https://aclanthology.org/2025.acl-long.572/",
    pages = "11663--11679",
    ISBN = "979-8-89176-251-0"
}

@inproceedings{ajith2024litsearch,
    title = "{L}it{S}earch: A Retrieval Benchmark for Scientific Literature Search",
    author = "Ajith, Anirudh  and
      Xia, Mengzhou  and
      Chevalier, Alexis  and
      Goyal, Tanya  and
      Chen, Danqi  and
      Gao, Tianyu",
    editor = "Al-Onaizan, Yaser  and
      Bansal, Mohit  and
      Chen, Yun-Nung",
    booktitle = "Proceedings of the 2024 Conference on Empirical Methods in Natural Language Processing",
    month = nov,
    year = "2024",
    address = "Miami, Florida, USA",
    publisher = "Association for Computational Linguistics",
    url = "https://aclanthology.org/2024.emnlp-main.840/",
    doi = "10.18653/v1/2024.emnlp-main.840",
    pages = "15068--15083"
}

@inproceedings{kang2024taxonomy,
    title = "Taxonomy-guided Semantic Indexing for Academic Paper Search",
    author = "Kang, SeongKu  and
      Zhang, Yunyi  and
      Jiang, Pengcheng  and
      Lee, Dongha  and
      Han, Jiawei  and
      Yu, Hwanjo",
    editor = "Al-Onaizan, Yaser  and
      Bansal, Mohit  and
      Chen, Yun-Nung",
    booktitle = "Proceedings of the 2024 Conference on Empirical Methods in Natural Language Processing",
    month = nov,
    year = "2024",
    address = "Miami, Florida, USA",
    publisher = "Association for Computational Linguistics",
    url = "https://aclanthology.org/2024.emnlp-main.407/",
    doi = "10.18653/v1/2024.emnlp-main.407",
    pages = "7169--7184"
}

@inproceedings{
mysore2021csfcube,
title={{CSFC}ube - A Test Collection of Computer Science Research Articles for Faceted Query by Example},
author={Sheshera Mysore and Tim O'Gorman and Andrew McCallum and Hamed Zamani},
booktitle={Thirty-fifth Conference on Neural Information Processing Systems Datasets and Benchmarks Track (Round 2)},
year={2021},
url={https://openreview.net/forum?id=8Y50dBbmGU}
}
